\documentclass[twocolumn]{aastex631}

\shorttitle{ H$\alpha$ and UV Probe of High-Redshift Outflows}
\shortauthors{Kehoe et al.}
\graphicspath{{./}}

\begin{document}

\title{The First Combined H$\alpha$ and Rest-UV Spectroscopic Probe of Galactic Outflows at High Redshift}

\author{Emily Kehoe}
\affiliation{Department of Physics \& Astronomy;
University of California, Los Angeles;
Los Angeles, CA 90095, USA}

\author{Alice E. Shapley}
\affiliation{Department of Physics \& Astronomy;
University of California, Los Angeles;
Los Angeles, CA 90095, USA}

\author{N.M. F\"{o}rster Schreiber}
\affiliation{Max-Planck-Institut f\"{u}r Extraterrestiche Physik (MPE);
Giessenbachstr., 85748, Garching, Germany}

\author{Anthony J. Pahl}
\affiliation{The Observatories of the Carnegie Institution for Science;
813 Santa Barbara Street;
Pasadena, CA 91101, USA}

\author{Michael W. Topping}
\affiliation{Department of Astronomy / Steward Observatory;
University of Arizona;
933 N Cherry Ave;
Tucson, AZ 85721, USA}

\author{Naveen A. Reddy}
\affiliation{Department of Physics \& Astronomy;
University of California, Riverside;
 900 University Avenue;
 Riverside, CA 92521, USA}

\author{Reinhard Genzel}
\affiliation{Max-Planck-Institut f\"{u}r Extraterrestiche Physik (MPE);
Giessenbachstr., 85748, Garching, Germany}

\author{Sedona H. Price}
\affiliation{ Department of Physics and Astronomy and PITT PACC;
University of Pittsburgh;
Pittsburgh, PA 15260, USA}

\author{L.J. Tacconi}
\affiliation{Max-Planck-Institut f\"{u}r Extraterrestiche Physik (MPE);
Giessenbachstr., 85748, Garching, Germany}



\begin{abstract}

We investigate the multi-phase structure of gas flows in galaxies. We study 80 galaxies during the epoch of peak star formation ($1.4\leq z\leq2.7$) using data from Keck/LRIS and VLT/KMOS. Our analysis provides a simultaneous probe of outflows using UV emission and absorption features and H$\alpha$ emission. With this unprecedented data set, we examine the properties of gas flows estimated from LRIS and KMOS in relation to other galaxy properties, such as star formation rate (SFR), star formation rate surface density ($\Sigma\textsubscript{SFR}$), stellar mass (M$_*$), and main sequence offset ($\Delta$MS). We find no strong correlations between outflow velocity measured from rest-UV lines centroids and galaxy properties. However, we find that galaxies with detected outflows show higher averages in SFR, $\Sigma\textsubscript{SFR}$, and $\Delta$MS than those lacking outflow detections, indicating a connection between outflow and galaxy properties. Furthermore, we find a lower average outflow velocity than previously reported, suggesting greater absorption at the systemic redshift of the galaxy. Finally, we detect outflows in 49\% of our LRIS sample and 30\% in the KMOS sample, and find no significant correlation between outflow detection and inclination. These results may indicate that outflows are not collimated and that H$\alpha$ outflows have a lower covering fraction than low-ionization interstellar absorption lines. Additionally, these tracers may be sensitive to different physical scales of outflow activity. A larger sample size with a wider dynamic range in galaxy properties is needed to further test this picture.

\end{abstract}



\section{Introduction}
Galaxy outflows play a crucial role in galaxy evolution over cosmic time, significantly impacting both galaxies and the intergalactic medium (IGM). Galaxy outflows, powered by phenomena such as supernovae, stellar winds, and active galactic nuclei (AGN), serve to regulate the availability of gas for star formation \citep{2001Heckman, 2006Croton}. The relationship between galaxy stellar mass or luminosity and dark matter halo mass is influenced by gas outflows \citep{1996Madau, 2010Moster, 2013Behroozi}. Furthermore,  galactic outflows regulate the chemical enrichment histories of galaxies \citep{2012Dave, Hopkins2012, 2013Hirschmann, 2013Vogelsberger, 2017Chisholm}. Specifically, outflows deplete the amount of cold gas available for star formation and remove metals from galaxies \citep{2005Scannapieco, 2005NDiMatteo, 2006Croton, 2008Somerville, 2015Erb, 2017Beckman}, enriching the circumgalactic medium (CGM) and the IGM \citep{2004Tremonti, 2007Dalcanton, 2008Finlator, 2014Peeples, 2017Tumlinson}. In other words, the role played by gas outflows in galaxy formation is reflected in the form of the mass-metallicity relation \citep{2011Dave, 2017Calabro, 2018Sanders, 2021Fontanot, 2021Sanders} and the fundamental metallicity relation \citep{2010Mannucci, 2018Sanders, 2021Sanders}.

The properties of outflows at high redshift have been investigated with both rest-UV interstellar features, such as metal absorption lines and Ly$\alpha$ emission, and broad rest-optical nebular line emission, such as H$\alpha$, [N\,{\sc ii}], [S\,{\sc ii}], and [O\,{\sc iii}]. Blueshifted interstellar absorption yields information on the outflowing material that galaxies have been ejecting throughout time \citep{2003Shapley, 2009Weiner, 2010Steidel, 2012Kornei, 2012Talia, 2014Bordoloi, 2022Calabro, 2022Weldon}. In rest-frame optical emission line spectra, broad high velocity components trace denser outflowing material that is within a few kiloparsecs of the launching points of outflows \citep{2009Shapiro, 2011Genzel, 2014Genzel, 2012aNewman, 2014Newman, 2016Cano, 2014Schreiber, 2019Schreiber, 2015Brusa, 2015Cresci, 2016Harrison, 2017Leung, 2019Leung, 2019Davies, 2020Davies, 2019Freeman, 2019Swinbank, 2022Concas}. However, rest-UV absorption and rest-optical nebular emission features tend to provide different answers on the nature of galaxy outflows, such as the detection rate, 3D structure, kinematics, and mass loading factors \citep{2003Shapley,2010Steidel,2012Talia,2019Schreiber,2019Davies, 2022Calabro,2022Weldon}.  This discrepancy implies that outflows have a complex and multi-phase structure. 

Thus far, no study has investigated rest-optical and rest-UV probes of outflows simultaneously in the same galaxies among the general population at $z \sim 2$, which is vital for understanding the multi-phase structure of outflowing gas. In this study, we analyze a sample of 80 galaxies at $1.4 \leq z \leq 2.7$ using observations obtained from the Low Resolution Imagining Spectrometer (LRIS) \citep{1995Oke,2004Steidel} at Keck. At these redshifts, LRIS spectra cover Ly$\alpha$ emission and various low-ionization interstellar (LIS) absorption lines that probe wind kinematics  (i.e., Si\,{\sc ii} $\lambda 1260$, O\,{\sc i} $\lambda 1302$, Si\,{\sc ii} $\lambda 1304$, C\,{\sc ii} $\lambda 1334$, and Si\,{\sc ii} $\lambda 1527$) as well as the frequency of outflow detections. We draw our sample from the KMOS$\textsuperscript{3D}$ survey \citep{2015Wisnioski, 2019Wisnioski}, which uses the K-Band Multi Object Spectrograph (KMOS) at the Very Large Telescope (VLT) to spatially and spectrally resolve the H$\alpha + $[N\,{\sc ii}]$+$[S\,{\sc ii}] line emission of $z\sim0.7-2.7$ star-forming galaxies. As part of the KMOS$\textsuperscript{3D}$ survey, the demographics and properties of galactic-scale outflows were studied by \cite{2014Genzel} and \cite{2019Schreiber}. 

We use this unique data set to study the correlation between outflow detection and galaxy properties by looking at the frequency of outflows determined from LRIS and KMOS. One of the primary objectives of this paper is to look for any trends between outflow velocities and galaxy properties, such as inclination ($i$), star formation rate (SFR), SFR surface-density ($\Sigma\textsubscript{SFR}$), stellar mass (M$_*$), and main sequence offset ($\Delta$MS). Furthermore, based on the unique combination of both rest-UV interstellar features and H$\alpha$ emission probes of outflows, we aim to analyze the outflow kinematics and investigate the geometry of galactic outflows.

The outline of this paper is as follows: Section \ref{sec:Data} introduces our observations, data reduction, and the final sample used for our analysis. Section \ref{sec:Measurements} describes the methods for measuring galactic properties (i.e., outflow velocity, $i$, SFR, $\Sigma\textsubscript{SFR}$, M$_*$, and $\Delta$MS). Section \ref{sec:Results} presents the results of our analysis of the correlations between outflow properties inferred from both LRIS and KMOS observations. Section \ref{sec:Discussion} discusses the implications of our key results. Throughout this paper, we adopt a $\Lambda$CDM cosmology with $H_0 = 70 \text{ km s}^{-1}$, $\Omega_m = 0.3$, $\Omega_\Lambda = 0.7$ and the \cite{2003Chabrier} stellar initial mass function (IMF).

\section{Data}\label{sec:Data}
\subsection{KMOS\textsuperscript{3D} Survey}
The KMOS$^{\text{3D}}$ survey was conducted with the multi-IFU instrument KMOS at the VLT \citep{2015Wisnioski, 2019Wisnioski} and focused on galaxies selected from the 3D-HST catalog \citep{2016AMomcheva}. The survey observed H$\alpha$, [N\,{\sc ii}], and [S\,{\sc ii}] emission in the YJ, H, and K bands for galaxies spanning a redshift range of $z = 0.7$--$2.7$. Building on earlier work on outflows based on near-IR IFU
observations with SINFONI from the SINS/zC-SINF survey \citep{2009Shapiro,2011Genzel,2012aNewman, 2012bNewman,2014Schreiber} and the first-year sample from KMOS\textsuperscript{3D} \citep{2014Genzel}, \cite{2019Schreiber} exploited the completed KMOS\textsuperscript{3D} survey data supplemented with smaller sets from SINS/zC-SINF and slit spectroscopic campaigns \citep{2007Kriek, 2008Kriek, 2013Genzel, 2014Newman, 2015vanDokkum} to characterize outflow demographics and properties.  

The full sample of 599 galaxies (525 from KMOS\textsuperscript{3D}) was used to search for a broad outflow emission signature in H$\alpha +$[N\,{\sc ii}]$+$[S\,{\sc ii}] emission, evident as residual high-velocity wings underneath the star formation-dominated narrow component in "velocity-shifted" spectra.  The IFU data allow mapping of the velocity field derived from the emission line peak, which can then be used to align the spectra of individual spaxels across the galaxies to a common peak velocity. The aligned spectra are then
added together to create "velocity-shifted" spectra.  This
technique removes the line broadening caused by gravitational
motions (e.g., disk rotation), facilitating the identification
of high-velocity components from outflows. For cases with identified outflow signature, \cite{2019Schreiber} attributed the outflow driver to star formation or AGN primarily on the basis of whether an AGN was identified through independent indicators (see below for more detail) as well as on the basis of the narrow [N\,{\sc ii}]/H$\alpha$ ratio. In total, \cite{2019Schreiber} found that within the KMOS$\textsuperscript{3D}$ survey there are 190 out of 599 galaxies at $1.4 \leq z \leq 2.7$ with a broad-component outflow signature, yielding an outflow detection fraction of 32\%. Among galaxies with outflow signatures, there are 87 galaxies with SF-driven winds (46\%) and 103 galaxies with AGN-driven winds (54\%).

\subsection{KMOS-LRIS Observations}

\subsubsection{Sample Selection}
\begin{figure*}
    \centering
    \includegraphics[width=7in]{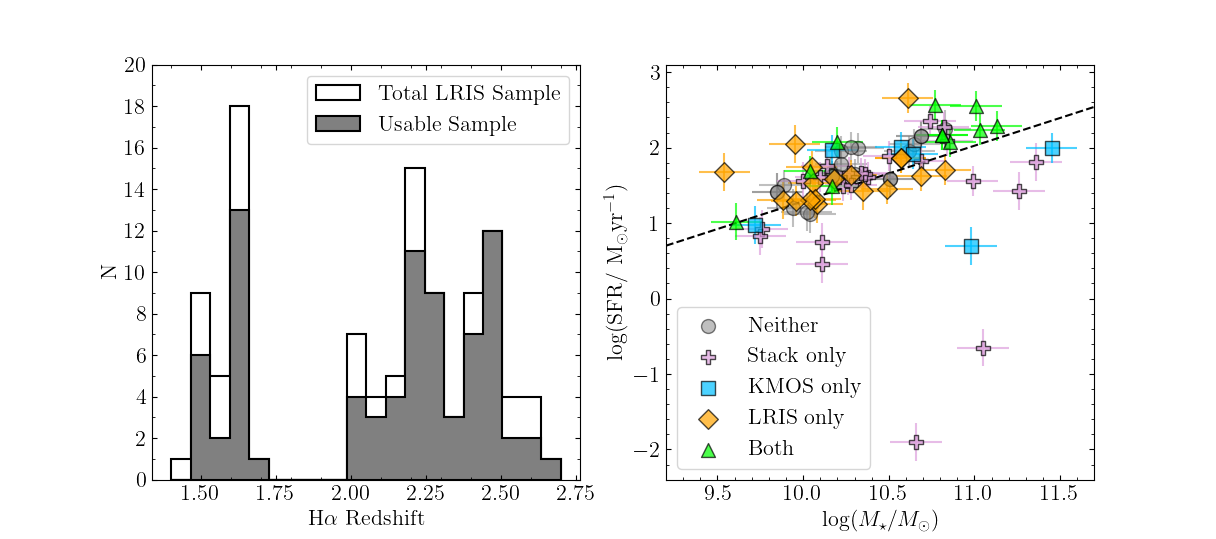}
    \caption{Properties of the full LRIS sample. \textit{Left:} H$\alpha$ redshift distributions for the LRIS sample. The open histogram represents the total sample (107 objects), and the gray histogram represents the sample of objects that had a usable spectrum for our analysis (80 objects). \textit{Right:} SFR versus stellar mass. SFRs and stellar masses are derived from \protect\cite{2019Schreiber} who followed procedures by \protect\cite{2011Wuyts}.  Gray circles are galaxies where outflows were not detected in either KMOS or LRIS (23 objects), purple crosses represent galaxies that were only used in creating stacked spectra because no significant line measurements were found (23 objects), blue squares are galaxies where outflows were detected in KMOS (6 objects), orange diamonds are galaxies where outflows were detected with only LRIS (17 objects), and green triangles are galaxies that where outflows were detected with both KMOS and LRIS (11 objects). LRIS outflows are identified in galaxies with $\Delta v_{\text{LIS}}$, $\Delta v_{\text{Ly}\alpha}$, or both significantly offset from the systemic velocity. The dashed line represents the SFR main sequence for $z\sim2$ from \cite{2014Speagle}. } 
    \label{fig:Ha_z_hist}
\end{figure*}

In constructing a sample for follow-up LRIS observations, we selected 85 galaxies from the KMOS$\textsuperscript{3D}$ survey \citep{2015Wisnioski}. These targets lie in the COSMOS and GOODS-S extragalactic legacy fields, and are covered by extensive existing multi-wavelength datasets. The mask design included all galaxies with KMOS emission line detections, prioritizing galaxies at $z>1.5$ with higher data quality (i.e., higher S/N and adaptive optics) and the presence of outflow signatures in KMOS detected by \cite{2019Schreiber}. Out of the 85 galaxies targeted with LRIS, 33 (39\%) were identified as having outflows from the KMOS$\textsuperscript{3D}$ survey. Specifically, 19 (22\%) of the targets have SF-driven outflows, 14 (16\%) of the targets have AGN-driven outflows, and 52 (61\%) of the targets had no outflow detection. As shown in Figure \ref{fig:Ha_z_hist}, our final sample spans a redshift range of 1.50 $\leq$ z $\leq$ 2.68.  

\subsubsection{Observations}
We observed 85 KMOS$\textsuperscript{3D}$ galaxies using the Keck Low-Resolution Imaging Spectrometer (LRIS) \citep{1995Oke,2004Steidel} over six and a half nights, including 4 nights in December of 2019 and 2.5 nights in January of 2021.  Our observations use four multi-object slit masks: two in the COSMOS and two in the GOODS-S fields. All masks used 1.2" slits. We employed the d500 dichroic for the December 2019 run with the 400 lines mm$^{-1}$ grism blazed at 3400 \r{A} on the blue side and the 600 lines mm$^{-1}$ grating blazed at 5000 \r{A} on the red side. In January 2021, due to red-side instrument problems, we only collected data on the blue side. For these observations, we used the d680 dichroic and the 400 lines mm$^{-1}$ grism. In our analysis we only use the blue side data, as it covers the rest-UV features of interest for all of our target galaxies. LRIS blue-side spectra yielded a resolution of R $\sim$  800. With this configuration, we have continuous wavelength coverage from the 3100 \r{A} atmospheric cut-off up to the d500 dichroic cut-off at 5000 \r{A} for 43 spectra taken during the December 2019 run. The remaining spectra taken with the d680 dichroic have varying red wavelength cut-offs that range from  5200-7650 \r{A} depending on the horizontal location of the slit on the multi-slit mask. Exposure times ranged from 5 to 19 hours. Weather conditions during the December 2019 run were poor, permitting data collection during only 1.5 of the 4 scheduled nights. When it was at least partially clear, seeing ranged from 0.5 to 1.3 arcsec. Conditions were clear throughout the 2.5 nights of the January 2021 run, with seeing ranging from 0.65 to 1.1 arcsec.  A summary of the masks used during our LRIS observations is provided in Table \ref{tab:masks}.

\begin{deluxetable*}{ccccccc}
    \tablecaption{Summary of LRIS observations         \label{tab:masks}}
    \tablehead{
    \colhead{Field}   & \colhead{Mask Name} & \colhead{RA}           & \colhead{Dec.}          & \colhead{t$\textsubscript{Blue}\textsuperscript{exp}$ (s)} & \colhead{N${\textsubscript{{targets}}}$\tablenotemark{a}} & \colhead{N${\textsubscript{success}}$\tablenotemark{b}}
    }
    \startdata
            COSMOS  & co\_kl1   & 10:00:24.620 & +02:15:13.078 &    76 540      & 23  & 19           \\
            COSMOS  & co\_kl2   & 10:00:29.041 & +02:24:24.528 &     41 400     & 20  & 13          \\
            GOODS-S & gs\_kl1   & 03:32:30.639 & -27:48:24.384 &    46 200      & 20  & 9          \\
            GOODS-S & gs-kl2    & 03:32:28.728 & -27:43:29.212 &     17 400     & 22  &         17 \\ 
            \enddata
        \tablenotetext{a}{Total number of targets on each mask.}
        \tablenotetext{b}{Total number of successful extractions.}
\end{deluxetable*}

\subsubsection{Data Reduction}
Blue-side LRIS data reduction was performed using custom IRAF, IDL, and Python scripts. The first step was to rectify each spectrum by fitting a polynomial to each 2D slit and transforming them to be rectangular. Next, we flat-fielded each exposure and cut out the slits for each object. Additionally, we background subtracted each object by removing cosmic rays and creating bad-pixel maps. We then created a summed mosaic of the 2D spectra, including science, arc and sky images, by first calculating the offsets between the individual science exposures, shifting, and averaging them. We performed a secondary background subtraction to avoid overestimating the background. After the initial background subtraction, many of the spectra sit in troughs with negative fluxes on each side of the spectrum. This trough arises from fitting a polynomial to the light across the entire slit. The light from the object biases the fit, overestimating the background. The second pass background subtraction excludes the object from the fit, removing any bias. To further ensure the accuracy of the background subtraction, we identified the locations of bright emission lines within the 2D spectra to mask any object traces and broaden the region around the emission lines, since the object trace is wider in these regions. Applying the newly defined masks prevents oversubtraction in these specific regions. Using these second pass images, we fit a line or a polynomial of order 2 to the science trace of each exposure to extract 1D spectra from our stacked 2D spectra. We applied the same extraction aperture to the arc and sky spectra. We then used the arc spectrum to determine the wavelength solution, which we applied to the sky spectrum. Following this step, we used the wavelengths of the bright sky lines to determine any required zero point shift in the wavelength solution. Finally, we flux calibrated our data using standard star observations. 

Most of our masks were collected during one epoch. However, the co\_kl1 mask data was collected during both the December 2019 and January 2021 runs. The reductions for each run were done separately up to the 1D extractions. We combined the 1D spectra of each galaxy using a S/N-weighted average. As listed in Table \ref{tab:masks}, we successfully extracted 58 usable spectra (N$\textsubscript{success}$) out of our 85 targets (N$\textsubscript{targets}$). The remaining 27 spectra were either too noisy or had artifacts, making them unusable for our analysis.

\subsubsection{MOSDEF-LRIS Observations}
We expanded our sample of KMOS$\textsuperscript{3D}$ galaxies with LRIS observations by incorporating observations from the MOSDEF-LRIS survey, described in detail by \cite{2020Topping}, \cite{2022Reddy}, and \cite{2022Weldon}. These MOSDEF-LRIS galaxies have KMOS$\textsuperscript{3D}$ coverage and fall within the same redshift range as the rest of the KMOS-LRIS sample (i.e., they satisfy the same selection criteria as the galaxies for which we obtained new LRIS observations). This LRIS survey targeted galaxies with existing MOSFIRE observations from the MOSFIRE Deep Evolution Field (MOSDEF) survey \citep{2015Kriek} and includes LRIS spectra with similar depth and the same observational setup as data we collected. The MOSDEF-LRIS and KMOS$\textsuperscript{3D}$ surveys both target the COSMOS and GOODS-S fields and have 22 galaxies in common. We integrate these observations into this paper to expand our KMOS$\textsuperscript{3D}$ sample with usable LRIS spectra from 58 to 80 galaxies. 

\section{Measurements}\label{sec:Measurements}
\subsection{Outflow Velocities} \label{sec:Outflow velocities}

\begin{deluxetable*}{cccc}
    \tablecaption{Spectral windows for line fitting      \label{tab:spec_windows}}
    \tablehead{
      \colhead{Line\tablenotemark{a}} & \colhead{$\lambda\textsubscript{rest}$ (\AA)} & \colhead{Blue Window (\AA)\tablenotemark{b}} & \colhead{Red Window (\AA)\tablenotemark{b}}
      }
      \startdata
      Ly$\alpha$ & 1215.67 & 1$195-1205$ & $1229-1234$\\
      Si\,{\sc ii} & 1260.42 & $1249-1235$ & $1280-1286$ \\
      O\,{\sc i} $+$ Si\,{\sc ii} & 1303.27 & $1286-1291$ & $1317-1323$ \\
      C\,{\sc ii} & 1334.53 & $1323-1329$ & $1348-1354$ \\
      Si\,{\sc ii} & 1526.71 & $1511-1517$ & $1560-1566$ \\
      \enddata
      \tablenotetext{a}{O\,{\sc i} $\lambda$1302 $+$ Si\,{\sc ii} $\lambda$1304 are blended at the resolution of our LRIS spectra.}
      \tablenotetext{b}{The blue and red windows are the wavelength intervals over which local continuum fitting was performed for each feature.}
\end{deluxetable*}

Large-scale gas outflows cause Doppler shifts in low-ionization interstellar (LIS) absorption lines and Ly$\alpha$ emission relative to the galaxy's systemic velocity. To quantify these shifts, we measure the line centroid velocity shifts, $\Delta v_{\text{LIS}}$ and $\Delta v_{\text{Ly}\alpha}$, from the centroid wavelengths, respectively, of strong LIS absorption lines and Ly$\alpha$ emission.  We first determined which lines were significantly-detected by using a non-parametric estimate of the line flux, which was used to find the equivalent width (EW) of each line. When determining the EW, we defined the continuum using blue and red wavelength windows around each line. These windows were defined using a stacked spectrum of the entire sample, ensuring a high S/N to provide a precise average for where the continuum lies. The wavelength windows are listed in Table \ref{tab:spec_windows}.  Uncertainties on the flux measurements were determined using Monte Carlo simulations in which each spectrum was perturbed on a pixel-by-pixel basis according to the error spectrum over 1000 iterations. If the absolute values of line fluxes were greater than 2$\sigma$, we labeled the line as being significantly-detected. For each significantly-detected line, we measured the observed centroids for the LIS absorption lines and the Ly$\alpha$ emission line, $\lambda_{\text{LIS, obs}}$ and $\lambda_{\text{Ly}\alpha \text{, obs}}$, respectively, by fitting Gaussian curves to the lines. Uncertainties for the centroid measurements were found using the same Monte Carlo simulations. We shifted $\lambda\textsubscript{LIS, obs}$ and $\lambda_{\text{Ly}\alpha\text{, obs}}$ to the rest frame defined by the object's H$\alpha$ redshift measured in the KMOS$\textsuperscript{3D}$ survey by \cite{2019Schreiber} (i.e., $\lambda\textsubscript{LIS} = \lambda\textsubscript{LIS, obs}/(1+z_{\text{H}\alpha})$ and $\lambda_{\text{Ly}\alpha} = \lambda_{\text{Ly}\alpha\text{, obs}}/(1+z_{\text{H}\alpha})$). We use these centroids to measure the line centroid velocity shifts: 
	\begin{equation}
	\Delta v_{\text{LIS}} = \frac{\lambda_{\text{LIS}} - \lambda_{\text{LIS, rest}}}{\lambda_{\text{LIS, rest}}}\times\text{c}
\end{equation} \begin{equation}
	\Delta v_{\text{Ly}\alpha} = \frac{\lambda_{\text{Ly}\alpha} - \lambda_{\text{Ly}\alpha\text{, rest}}}{\lambda_{\text{Ly}\alpha\text{, rest}}}\times\text{c}
\end{equation} 
where $\lambda\textsubscript{LIS, rest}$ and $\lambda_{\text{Ly}\alpha\text{, rest}}$ are the laboratory wavelengths for these features as listed in Table \ref{tab:spec_windows}. From our total sample of 80 usable galaxies, 57 (67\%) were found to have at least one significant rest-UV line measurement in the LRIS spectra. To find $\Delta v_{\text{LIS}}$, we calculated an inverse-variance-weighted average of the individual $\Delta v_{\text{LIS}}$ measurements for every object that had at least one significantly-detected LIS line. The LIS lines used in the weighted average were Si\,{\sc ii}$\lambda 1260$, C\,{\sc ii}$\lambda 1334$, and Si\,{\sc ii}$\lambda 1526$. We do not include the blend of O\,{\sc ii}$\lambda 1302 +$ Si\,{\sc ii}$\lambda 1304$ in our measurements because the blend centroid wavelength is harder to constrain. Velocities that were greater in magntiude than 1$\sigma$ from zero were defined as having a significant flow. Based on the error bars presented in Table \ref{tab:outflow_detections}, the minimum outflow velocity we can detect is $15-20$ km/s in the highest S/N cases.

\begin{figure}
    \centering
    \includegraphics[width=\linewidth]{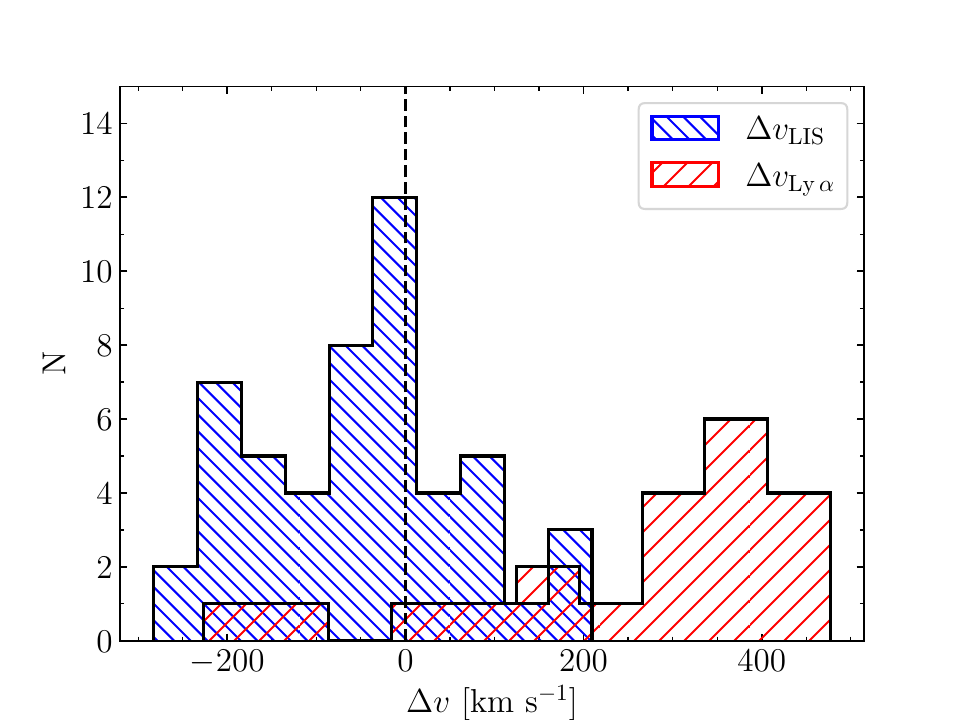}
    \caption{Velocity offset distribution. The blue hashed histogram represents centroid velocity shifts for LIS absorption lines and the red hashed histogram represents centroid velocity shifts for Ly$\alpha$ emission. In total, there are 57 objects with at least one feature. 51 objects had LIS velocity measurements with $\langle \Delta v_{\text{LIS}}\rangle= -56 \pm 16 \text{km s}^{-1}$ and 21 objects had Ly$\alpha$ velocity measurements with $\langle \Delta v_{\text{Ly}\alpha} \rangle =  266 \pm 41 \text{km s}^{-1}$.}
    \label{fig:Delta_v_hist}
\end{figure}

Out of the 57 significant LIS line measurements, 19 were found to have a significantly-detected outflow (i.e., negative velocity) and 6 were found to have a significantly-detected inflow (i.e., positive velocity). In terms of Ly$\alpha$ kinematics, 17 objects had a significantly-detected outflow (i.e., positive velocity) and 2 objects had a significantly-detected inflow (i.e., negative velocity). Specifically, 11 galaxies show a significant outflow detected solely from LIS absorption lines, 9 galaxies have significant outflows detected exclusively from Ly$\alpha$ emission, and 9 galaxies have significant outflows detected in both. Outflows may only be detected from only Ly$\alpha$ emission because galaxies may have weak absorption lines, but strong Ly$\alpha$ emission. Conversely, galaxies detected by only LIS absorption may show only strong broad Ly$\alpha$ absorption \citep{2003Shapley}. Indeed, the Ly$\alpha$ profile is complex and can range from well-detected emission to strong, broad absorption. In the case of such strong absorption, no meaningful $\Delta v\textsubscript{Ly$\alpha$}$  can be measured. Figure \ref{fig:Delta_v_hist} shows the distribution of our velocity offset measurements. The mean $\Delta v\textsubscript{LIS}$ for our sample is $-56 \pm 16$ km s$^{-1}$ and the mean $\Delta v_{\text{Ly}\alpha}$ is $+266 \pm$ 41 km s$^{-1}$. The magnitudes of the $\Delta v\textsubscript{LIS}$ and $\Delta v\textsubscript{Ly$\alpha$}$ are not strongly correlated.

We do not present outflow velocities from H$\alpha$ measurements because the FWHM as a measure of outflow velocity differs from the centroid shifts used for LIS and Ly$\alpha$ in LRIS measurements. The velocity offsets based on centroids of broad components from KMOS$\textsuperscript{3D}$ stacked spectra are generally modest and poorly constrained, except for stronger outflows. Accordingly, comparing kinematic measurements from the two samples is nontrivial. Additionally, emission and absorption lines trace material differently, with H$\alpha$ more sensitive to denser gas due to its electron density dependence \citep{2019Schreiber}, typically tracing material closer to the galaxy compared to the more extended regions traced by LIS.

As shown in Table \ref{tab:outflow_detections}, 57 galaxies had at least one feature measured with the LRIS spectra. Out of this sample, 17 galaxies had a significant outflow detected with LRIS only, 6 galaxies had a significant outflow detected with KMOS only, 11 had significant detections with both LRIS and KMOS, and 23 had a detection from neither LRIS or KMOS. The detection fraction of outflows with LRIS and KMOS was 49\% (28 out of 57 galaxies) and 30\% (17 out of 57 galaxies), respectively.

Furthermore, we measure the maximum outflow velocity ($v$\textsubscript{max}) by using methods described by \cite{2012Kornei} and \cite{2022Weldon}. In summary, we determine the minimum of each LIS absorption line and evaluate the sum of the flux and its uncertainty at each wavelength increment. The spectrum is evaluated at shorter wavelengths until the sum of flux plus uncertainty surpasses 1.0. The first wavelength at which this occurs is the wavelength used to calculate the $v$\textsubscript{max}. Uncertainties are found using the same Monte Carlo simulations previously described.

\begin{deluxetable*}{cccccccccccc}
  \centering
    \tabletypesize{\scriptsize}
    \tablecaption{Measurements and detections for galactic outflows       \label{tab:outflow_detections}}
    \tablehead{
    \colhead{Object} & \colhead{RA} & \colhead{Dec} & \colhead{$z_{\text{H}\alpha}$} & \colhead{{$\Delta v_{\text{LIS}}$}\tablenotemark{a}} & \colhead{{$\Delta v_{\text{Lya}}$}\tablenotemark{a}} & \colhead{{$\Delta v_{\text{max}}$}\tablenotemark{a}} & \colhead{LRIS Outflow\tablenotemark{b}} & \colhead{KMOS Outflows\tablenotemark{c}} & \colhead{AGN\tablenotemark{d}} \\ & & & & \colhead{$[\text{km s}^{-1}]$} & \colhead{$[\text{km s}^{-1}]$} & \colhead{$[\text{km s}^{-1}]$} & & &
    }
    \startdata
COS4\_12476 & 10:00:27.638 & +02:18:24.773 & 1.5143 &  $-$130 $\pm$43 &           ... & $-$187 $\pm$109 &        1 &        0 &    0 \\
COS4\_24738 & 10:00:33.201 & +02:26:02.811 & 1.5888 &    19 $\pm$94 &           ... & $-$287 $\pm$208 &        0 &        0 &    0 \\
COS4\_12056 & 10:00:31.208 & +02:18:09.725 & 1.6000 &  113 $\pm$282 &           ... & $-$225 $\pm$197 &        0 &        1 &    0 \\
 GS4\_39085 & 03:32:17.113 & $-$27:43:42.067 & 1.6100 &            ... & 402 $\pm$526 & ... &        1 &        0 &    0 \\
 GS4\_08422 & 03:32:37.761 & $-$27:52:12.306 & 1.6113 &   $-$33 $\pm$42 &           ... &  $-$87 $\pm$282 &        0 &        1 &    1 \\
 GS4\_44066 & 03:32:25.165 & $-$27:42:18.785 & 1.6140 &            ... & 476 $\pm$259 & ... &        1 &        1 &    1 \\
 GS4\_11203 & 03:32:36.206 & $-$27:51:29.923 & 1.6144 &   $-$98 $\pm$80 &           ... & $-$285 $\pm$153 &        1 &        0 &    0 \\
COS4\_11343 & 10:00:35.251 & +02:17:43.035 & 1.6474 &    13 $\pm$29 &  $-$98 $\pm$54 &  $-$346 $\pm$86 &        0 &        0 &    1 \\
COS4\_18358 & 10:00:40.111 & +02:22:00.462 & 1.6484 &   $-$20 $\pm$49 &  399 $\pm$19 &  $-$75 $\pm$107 &        1 &        1 &    0 \\
COS4\_20595 & 10:00:39.360 & +02:23:20.651 & 1.6547 &  $-$183 $\pm$19 &  444 $\pm$26 &  $-$118 $\pm$74 &        1 &        0 &    0 \\
COS4\_20449 & 10:00:28.246 & +02:23:15.611 & 1.6559 &   $-$46 $\pm$50 &           ... &  $-$300 $\pm$87 &        0 &        0 &    0 \\
COS4\_17519 & 10:00:36.870 & +02:21:30.183 & 1.7081 &  105 $\pm$152 &           ... & $-$220 $\pm$158 &        0 &        0 &    0 \\
COS4\_18604 & 10:00:31.758 & +02:22:08.159 & 2.0055 &  $-$162 $\pm$89 &           ... &  $-$518 $\pm$99 &        1 &        0 &    0 \\
COS4\_20746 & 10:00:38.767 & +02:23:27.429 & 2.0070 &  $-$282 $\pm$84 &           ... &  $-$445 $\pm$73 &        1 &        0 &    0 \\
 GS4\_20410 & 03:32:21.950 & $-$27:48:55.602 & 2.0085 &   19 $\pm$100 &           ... & $-$158 $\pm$165 &        0 &        0 &    0 \\
COS4\_13174 & 10:00:26.935 & +02:18:50.313 & 2.0974 & $-$278 $\pm$184 &  270 $\pm$25 & $-$396 $\pm$224 &        1 &        1 &    0 \\
 GS4\_42363 & 03:32:28.410 & $-$27:42:46.562 & 2.1408 &  $-$218 $\pm$26 &           ... &  $-$563 $\pm$51 &        1 &        1 &    0 \\
 GS4\_41886 & 03:32:23.436 & $-$27:42:55.015 & 2.1411 &            ... &   271 $\pm$9 & ... &        1 &        1 &    1 \\
COS4\_08775 & 10:00:16.549 & +02:16:09.402 & 2.1624 &  $-$122 $\pm$84 &  372 $\pm$34 & $-$287 $\pm$162 &        1 &        0 &    0 \\
COS4\_13701 & 10:00:27.052 & +02:19:09.982 & 2.1664 &   $-$41 $\pm$51 &           ... & $-$113 $\pm$137 &        0 &        1 &    0 \\
COS4\_25229 & 10:00:26.019 & +02:26:22.974 & 2.1807 &   $-$72 $\pm$17 &           ... &  $-$430 $\pm$41 &        1 &        0 &    0 \\
 GS4\_38116 & 03:32:41.113 & $-$27:43:58.606 & 2.1966 &  $-$26 $\pm$131 &           ... &            ... &        0 &        0 &    0 \\
 GS4\_38116 & 03:32:41.113 & $-$27:43:58.606 & 2.1966 &  $-$30 $\pm$116 &           ... &            ... &        0 &        0 &    0 \\
COS4\_09044 & 10:00:35.706 & +02:16:19.384 & 2.1983 &   $-$16 $\pm$44 &           ... & $-$191 $\pm$131 &        0 &        0 &    0 \\
 GS4\_25151 & 03:32:23.914 & $-$27:47:39.386 & 2.2229 &   175 $\pm$81 &           ... & $-$240 $\pm$122 &        0 &        0 &    0 \\
 GS4\_29868 & 03:32:29.066 & $-$27:46:28.614 & 2.2239 &   $-$29 $\pm$56 &           ... & $-$371 $\pm$107 &        0 &        0 &    0 \\
COS4\_04930 & 10:00:29.037 & +02:13:43.661 & 2.2273 &    66 $\pm$77 &           ... & $-$731 $\pm$132 &        0 &        0 &    0 \\
COS4\_04930 & 10:00:29.037 & +02:13:43.661 & 2.2273 &    66 $\pm$79 &           ... & $-$69 $\pm$194 &        0 &        0 &    0 \\
COS4\_04519 & 10:00:28.641 & +02:13:26.952 & 2.2285 &  $-$185 $\pm$80 &  223 $\pm$98 &  $-$948 $\pm$96 &        1 &        0 &    0 \\
COS4\_06963 & 10:00:18.380 & +02:14:58.858 & 2.3012 & $-$213 $\pm$216 & $-$227 $\pm$59 & $-$148 $\pm$451 &        0 &        1 &    0 \\
 GS4\_41748 & 03:32:24.196 & $-$27:42:57.553 & 2.3013 &            ... &  109 $\pm$54 & ... &        1 &        0 &    1 \\
COS4\_05389 & 10:00:17.593 & +02:13:58.786 & 2.3013 &   $-$42 $\pm$97 &           ... & $-$399 $\pm$167 &        0 &        1 &    0 \\
 GS4\_40768 & 03:32:09.797 & $-$27:43:08.645 & 2.3033 &    82 $\pm$19 &           ... &  $-$290 $\pm$43 &        0 &        0 &    1 \\
 GS4\_36705 & 03:32:10.189 & $-$27:44:16.303 & 2.3055 &   $-$54 $\pm$26 &  378 $\pm$18 & $-$152 $\pm$118 &        1 &        0 &    0 \\
COS4\_01966 & 10:00:30.209 & +02:11:57.563 & 2.3058 &   $-$1 $\pm$201 &           ... & $-$295 $\pm$261 &        0 &        0 &    0 \\
COS4\_03324 & 10:00:35.618 & +02:12:47.281 & 2.3069 &  $-$65 $\pm$118 &           ... & $-$112 $\pm$219 &        0 &        0 &    0 \\
COS4\_02672 & 10:00:31.073 & +02:12:25.912 & 2.3077 &  $-$212 $\pm$50 &           ... &  $-$634 $\pm$96 &        1 &        0 &    0 \\
COS4\_02672 & 10:00:31.073 & +02:12:25.912 & 2.3077 &   $-$80 $\pm$73 &  328 $\pm$66 &  $-$521 $\pm$241 &        1 &        0 &    0 \\
 GS4\_38807 & 03:32:43.633 & $-$27:43:47.712 & 2.3177 & $-$215 $\pm$263 &           ... & $-$258 $\pm$261 &        0 &        0 &    0 \\
 GS4\_35937 & 03:32:38.139 & $-$27:44:33.630 & 2.3292 & $-$224 $\pm$124 &           ... & $-$137 $\pm$238 &        1 &        0 &    1 \\
 GS4\_46938 & 03:32:32.294 & $-$27:41:26.362 & 2.3323 &   $-$80 $\pm$15 &   294 $\pm$8 &   $-$40 $\pm$73 &        1 &        1 &    0 \\
 GS4\_45188 & 03:32:15.182 & $-$27:41:58.693 & 2.4061 & $-$186 $\pm$103 &  146 $\pm$40 & $-$225 $\pm$379 &        1 &        1 &    1 \\
 GS4\_45188 & 03:32:15.182 & $-$27:41:58.693 & 2.4061 &  $-$35 $\pm$274 & 374 $\pm$103 & $-$924 $\pm$168 &        1 &        1 &    1 \\
 GS4\_40679 & 03:32:19.057 & $-$27:43:15.143 & 2.4079 &   209 $\pm$67 &           ... & $-$508 $\pm$374 &        0 &        0 &    0 \\
 GS4\_40679 & 03:32:19.057 & $-$27:43:15.143 & 2.4079 &   209 $\pm$63 &           ... & $-$157 $\pm$115 &        0 &        0 &    0 \\
 GS4\_38560 & 03:32:18.726 & $-$27:43:51.672 & 2.4165 & $-$146 $\pm$134 &           ... & $-$396 $\pm$216 &        1 &        0 &    0 \\
COS4\_06079 & 10:00:26.272 & +02:14:24.258 & 2.4413 &            ... &   25 $\pm$55 & ... &        0 &        1 &    0 \\
COS4\_17298 & 10:00:32.355 & +02:21:21.002 & 2.4443 &   $-$98 $\pm$29 &  430 $\pm$15 &  $-$442 $\pm$55 &        1 &        0 &    0 \\
 GS4\_40218 & 03:32:38.869 & $-$27:43:21.565 & 2.4504 &    62 $\pm$28 &           ... & $-$235 $\pm$224 &        0 &        0 &    0 \\
 GS4\_40218 & 03:32:38.869 & $-$27:43:21.565 & 2.4504 &    27 $\pm$25 &           ... & $-$271 $\pm$67 &        0 &        0 &    0 \\
 GS4\_45068 & 03:32:33.016 & $-$27:42:00.378 & 2.4527 &            ... &  161 $\pm$18 & ... &        1 &        1 &    1 \\
COS4\_08515 & 10:00:44.275 & +02:15:58.544 & 2.4539 &    9 $\pm$105 &  429 $\pm$12 &  $-$31 $\pm$156 &        1 &        0 &    0 \\
COS4\_12148 & 10:00:28.499 & +02:18:09.696 & 2.4603 &     4 $\pm$61 &           ... & $-$262 $\pm$128 &        0 &        0 &    0 \\
COS4\_22995 & 10:00:17.153 & +02:24:52.319 & 2.4681 &  $-$34 $\pm$165 &  364 $\pm$43 & $-$657 $\pm$207 &        1 &        1 &    1 \\
COS4\_22564 & 10:00:17.563 & +02:24:42.596 & 2.4694 &   $-$26 $\pm$81 &           ... &  $-$562 $\pm$75 &        0 &        0 &    0 \\
COS4\_27120 & 10:00:24.075 & +02:27:45.211 & 2.4780 &  $-$171 $\pm$60 &           ... & $-$336 $\pm$112 &        1 &        1 &    0 \\
COS4\_27087 & 10:00:24.214 & +02:27:41.260 & 2.4794 &  $-$149 $\pm$37 &           ... &  $-$677 $\pm$67 &        1 &        0 &    0 \\
    \enddata
    \tablenotetext{a}{ "..." indicates that no significant ($>2\sigma$) detections of LIS or Ly$\alpha$ features were made in the LRIS spectrum.}
  \tablenotetext{b}{A significant LRIS outflow detection is denoted with 1, while a non-significant detection is denoted with 0. A detection is classified as significant when the measured $\Delta v\textsubscript{LIS}$  or $\Delta v_{\mathrm{Ly}\alpha}$ is greater in magnitude than 1$\sigma$ from 0.}
  \tablenotetext{c}{A KMOS outflow detection is denoted with 1, while a non-detection is denoted with 0. Detections are found from SF or AGN broad emission line signatures.}
  \tablenotetext{d}{Galaxies hosting AGN based on H$\alpha$ multi-wavelength analysis or rest-UV spectra are denoted with 1 and galaxies without AGN are denoted as 0. 10 out of the 15 AGNs in our sample have individual gas kinematic measurements.}
\end{deluxetable*}

\begin{deluxetable*}{ccccccc}
\caption{Galactic properties for the full KMOS-LRIS sample\label{tab:galaxy_properties}}
        \tablehead{
            \colhead{Object} &    \colhead{$z_{{\rm H}\alpha}$} &  \colhead{$i$ (degrees)} &  \colhead{log(SFR/(M$_{\odot}$yr$^{-1}$))} &  \colhead{$\log(\Sigma\textsubscript{SFR}$/(M$_{\odot} \text{yr}^{-1}\text{kpc}^{-2})$)} &  \colhead{$\log(\mathrm{M}_{*}/\text{M}_{\odot}$)} &   \colhead{$\Delta$MS} 
            }
        \startdata
        COS4\_12476 & 1.5143 & 59 $\pm$ 0.02 &  1.75  $\pm$  0.2 &   0.4  $\pm$  0.13 & 10.05  $\pm$  0.15 &    0.5  $\pm$  0.2 \\
        COS4\_24738 & 1.5888 & 49 $\pm$ 0.04 & 1.15  $\pm$  0.25 & -0.49  $\pm$  0.03 & 10.02  $\pm$  0.15 &  -0.1  $\pm$  0.25 \\
        COS4\_12056 & 1.6000 & 55 $\pm$ 0.03 & 0.98  $\pm$  0.25 &  -1.31  $\pm$  0.0 &  9.72  $\pm$  0.15 &  0.01  $\pm$  0.25 \\
         GS4\_39085 & 1.6100 & 44 $\pm$ 0.03 &  1.45  $\pm$  0.2 & -0.17  $\pm$  0.06 & 10.49  $\pm$  0.15 &  -0.15  $\pm$  0.2 \\
         GS4\_08422 & 1.6113 & 31 $\pm$ 0.03 &   2.0  $\pm$  0.2 &  1.62  $\pm$  1.34 & 11.45  $\pm$  0.15 &  -0.18  $\pm$  0.2 \\
         GS4\_44066 & 1.6140 & 27 $\pm$ 0.03 &  2.56  $\pm$  0.2 &  1.97  $\pm$  4.69 & 10.77  $\pm$  0.15 &   0.78  $\pm$  0.2 \\
         GS4\_11203 & 1.6144 & 45 $\pm$ 0.04 &  1.29  $\pm$  0.2 &  0.49  $\pm$  0.19 &  9.96  $\pm$  0.15 &   0.09  $\pm$  0.2 \\
        COS4\_11343 & 1.6474 & 53 $\pm$ 0.02 &  1.51  $\pm$  0.2 & -0.49  $\pm$  0.02 &  9.89  $\pm$  0.15 &   0.37  $\pm$  0.2 \\
        COS4\_18358 & 1.6484 & 53 $\pm$ 0.03 & 1.02  $\pm$  0.25 & -0.46  $\pm$  0.02 &  9.61  $\pm$  0.15 &  0.15  $\pm$  0.25 \\
        COS4\_20595 & 1.6547 & 58 $\pm$ 0.01 & 1.25  $\pm$  0.25 & -0.34  $\pm$  0.02 & 10.08  $\pm$  0.15 & -0.08  $\pm$  0.25 \\
        COS4\_20449 & 1.6559 & 34 $\pm$ 0.03 & 1.12  $\pm$  0.25 & -0.74  $\pm$  0.01 & 10.04  $\pm$  0.15 & -0.17  $\pm$  0.25 \\
        COS4\_17519 & 1.7081 & 39 $\pm$ 0.02 &   2.0  $\pm$  0.2 &  0.16  $\pm$  0.05 & 10.32  $\pm$  0.15 &   0.49  $\pm$  0.2 \\
        COS4\_18604 & 2.0055 & 64 $\pm$ 0.02 &  1.53  $\pm$  0.2 & -0.17  $\pm$  0.03 & 10.06  $\pm$  0.15 &   0.11  $\pm$  0.2 \\
        COS4\_20746 & 2.0070 & 32 $\pm$ 0.06 & 1.32  $\pm$  0.25 &  0.23  $\pm$  0.08 & 10.07  $\pm$  0.15 & -0.11  $\pm$  0.25 \\
         GS4\_20410 & 2.0085 & 56 $\pm$ 0.01 &  1.63  $\pm$  0.2 & -0.04  $\pm$  0.02 &  10.2  $\pm$  0.15 &   0.08  $\pm$  0.2 \\
        COS4\_13174 & 2.0974 & 73 $\pm$ 0.02 &  2.24  $\pm$  0.2 & -0.15  $\pm$  0.05 & 11.03  $\pm$  0.15 &    0.1  $\pm$  0.2 \\
         GS4\_42363 & 2.1408 & 68 $\pm$ 0.01 &  2.07  $\pm$  0.2 &  0.65  $\pm$  0.13 &  10.2  $\pm$  0.15 &   0.48  $\pm$  0.2 \\
         GS4\_41886 & 2.1411 & 54 $\pm$ 0.02 &  2.07  $\pm$  0.2 &  0.99  $\pm$  0.42 & 10.86  $\pm$  0.15 &   0.04  $\pm$  0.2 \\
        COS4\_08775 & 2.1624 & 55 $\pm$ 0.02 & 1.43  $\pm$  0.25 & -0.47  $\pm$  0.02 & 10.35  $\pm$  0.15 & -0.26  $\pm$  0.25 \\
        COS4\_13701 & 2.1664 & 50 $\pm$ 0.02 &  1.92  $\pm$  0.2 & -0.08  $\pm$  0.04 & 10.64  $\pm$  0.15 &   0.03  $\pm$  0.2 \\
        COS4\_25229 & 2.1807 & 48 $\pm$ 0.04 &  1.3  $\pm$  0.25 &  0.33  $\pm$  0.11 & 10.04  $\pm$  0.15 & -0.14  $\pm$  0.25 \\
         GS4\_38116 & 2.1966 & 64 $\pm$ 0.02 &  1.63  $\pm$  0.2 &   0.43  $\pm$  0.1 & 10.17  $\pm$  0.15 &   0.06  $\pm$  0.2 \\
         GS4\_38116 & 2.1966 & 64 $\pm$ 0.02 &  1.63  $\pm$  0.2 &   0.43  $\pm$  0.1 & 10.17  $\pm$  0.15 &   0.06  $\pm$  0.2 \\
        COS4\_09044 & 2.1983 & 57 $\pm$ 0.03 &  1.2  $\pm$  0.25 &  0.03  $\pm$  0.06 &  9.94  $\pm$  0.15 & -0.15  $\pm$  0.25 \\
         GS4\_25151 & 2.2229 & 28 $\pm$ 0.03 &  2.05  $\pm$  0.2 &  -0.2  $\pm$  0.02 & 10.65  $\pm$  0.15 &   0.13  $\pm$  0.2 \\
         GS4\_29868 & 2.2239 & 58 $\pm$ 0.02 &  2.01  $\pm$  0.2 &  -0.18  $\pm$  0.1 & 10.28  $\pm$  0.15 &   0.35  $\pm$  0.2 \\
        COS4\_04930 & 2.2273 & 59 $\pm$ 0.03 & 1.59  $\pm$  0.25 & -0.36  $\pm$  0.08 & 10.51  $\pm$  0.15 & -0.23  $\pm$  0.25 \\
        COS4\_04930 & 2.2273 & 59 $\pm$ 0.03 & 1.59  $\pm$  0.25 & -0.36  $\pm$  0.08 & 10.51  $\pm$  0.15 & -0.23  $\pm$  0.25 \\
        COS4\_04519 & 2.2285 & 50 $\pm$ 0.02 &  2.66  $\pm$  0.2 &  1.11  $\pm$  0.76 & 10.61  $\pm$  0.15 &   0.77  $\pm$  0.2 \\
        COS4\_06963 & 2.3012 & 28 $\pm$ 0.07 &  0.7  $\pm$  0.25 & -0.74  $\pm$  0.04 & 10.98  $\pm$  0.15 & -1.46  $\pm$  0.25 \\
        COS4\_05389 & 2.3013 & 47 $\pm$ 0.05 &  1.97  $\pm$  0.2 &  0.52  $\pm$  0.31 & 10.17  $\pm$  0.15 &   0.38  $\pm$  0.2 \\
         GS4\_41748 & 2.3013 & 36 $\pm$ 0.03 &   1.7  $\pm$  0.2 &  0.61  $\pm$  0.23 & 10.83  $\pm$  0.15 &  -0.35  $\pm$  0.2 \\
         GS4\_40768 & 2.3033 & 72 $\pm$ 0.01 &  1.78  $\pm$  0.2 & -0.07  $\pm$  0.01 & 10.22  $\pm$  0.15 &   0.14  $\pm$  0.2 \\
         GS4\_36705 & 2.3055 & 53 $\pm$ 0.02 &  1.64  $\pm$  0.2 &  0.41  $\pm$  0.16 & 10.28  $\pm$  0.15 &  -0.04  $\pm$  0.2 \\
        COS4\_01966 & 2.3058 & 65 $\pm$ 0.06 &  1.67  $\pm$  0.2 &  0.01  $\pm$  0.17 & 10.16  $\pm$  0.15 &   0.08  $\pm$  0.2 \\
        COS4\_03324 & 2.3069 & 46 $\pm$ 0.02 &  1.96  $\pm$  0.2 & -0.18  $\pm$  0.03 & 10.62  $\pm$  0.15 &   0.05  $\pm$  0.2 \\
        COS4\_02672 & 2.3077 & 60 $\pm$ 0.02 &  1.86  $\pm$  0.2 & -0.17  $\pm$  0.03 & 10.57  $\pm$  0.15 &  -0.02  $\pm$  0.2 \\
        COS4\_02672 & 2.3077 & 60 $\pm$ 0.02 &  1.86  $\pm$  0.2 & -0.17  $\pm$  0.04 & 10.57  $\pm$  0.15 &  -0.02  $\pm$  0.2 \\
         GS4\_38807 & 2.3177 & 64 $\pm$ 0.01 &  1.65  $\pm$  0.2 &  0.05  $\pm$  0.04 &  10.3  $\pm$  0.15 &  -0.04  $\pm$  0.2 \\
         GS4\_35937 & 2.3292 & 68 $\pm$ 0.01 &  1.62  $\pm$  0.2 & -0.61  $\pm$  0.02 & 10.69  $\pm$  0.15 &  -0.34  $\pm$  0.2 \\
         GS4\_46938 & 2.3323 & 58 $\pm$ 0.02 &  1.69  $\pm$  0.2 &  0.74  $\pm$  0.19 & 10.04  $\pm$  0.15 &   0.22  $\pm$  0.2 \\
         GS4\_45188 & 2.4061 & 25 $\pm$ 0.07 &  2.17  $\pm$  0.2 &  1.23  $\pm$  1.43 & 10.81  $\pm$  0.15 &    0.1  $\pm$  0.2 \\
         GS4\_45188 & 2.4061 & 25 $\pm$ 0.07 &  2.17  $\pm$  0.2 &  1.23  $\pm$  1.43 & 10.81  $\pm$  0.15 &    0.1  $\pm$  0.2 \\
         GS4\_40679 & 2.4079 & 36 $\pm$ 0.04 &  2.15  $\pm$  0.2 &   0.14  $\pm$  0.2 & 10.69  $\pm$  0.15 &   0.17  $\pm$  0.2 \\
         GS4\_40679 & 2.4079 & 36 $\pm$ 0.04 &  2.15  $\pm$  0.2 &  0.14  $\pm$  0.19 & 10.69  $\pm$  0.15 &   0.17  $\pm$  0.2 \\
         GS4\_38560 & 2.4165 & 49 $\pm$ 0.02 &  1.58  $\pm$  0.2 & -0.32  $\pm$  0.02 & 10.18  $\pm$  0.15 &  -0.05  $\pm$  0.2 \\
        COS4\_06079 & 2.4413 & 46 $\pm$ 0.03 &  2.01  $\pm$  0.2 & -0.23  $\pm$  0.03 & 10.57  $\pm$  0.15 &   0.11  $\pm$  0.2 \\
        COS4\_17298 & 2.4443 & 30 $\pm$ 0.07 & 1.68  $\pm$  0.25 &  0.37  $\pm$  0.13 &  9.54  $\pm$  0.15 &  0.67  $\pm$  0.25 \\
         GS4\_40218 & 2.4504 & 48 $\pm$ 0.03 & 1.41  $\pm$  0.25 &  0.23  $\pm$  0.07 &  9.85  $\pm$  0.15 &  0.09  $\pm$  0.25 \\
         GS4\_40218 & 2.4504 & 48 $\pm$ 0.03 & 1.41  $\pm$  0.25 &  0.23  $\pm$  0.07 &  9.85  $\pm$  0.15 &  0.09  $\pm$  0.25 \\
         GS4\_45068 & 2.4527 &  14 $\pm$ 0.1 &  2.55  $\pm$  0.2 &   1.66  $\pm$  3.3 & 11.01  $\pm$  0.15 &   0.35  $\pm$  0.2 \\
        COS4\_08515 & 2.4539 & 77 $\pm$ 0.02 &  1.3  $\pm$  0.25 &  -0.8  $\pm$  0.01 &  9.88  $\pm$  0.15 & -0.05  $\pm$  0.25 \\
        COS4\_12148 & 2.4603 & 49 $\pm$ 0.05 &  1.96  $\pm$  0.2 &  1.28  $\pm$  1.32 & 10.22  $\pm$  0.15 &   0.29  $\pm$  0.2 \\
        COS4\_22995 & 2.4681 & 48 $\pm$ 0.03 &  2.29  $\pm$  0.2 &  1.33  $\pm$  1.08 & 11.13  $\pm$  0.15 &    0.0  $\pm$  0.2 \\
        COS4\_22564 & 2.4694 & 47 $\pm$ 0.02 & 2.25  $\pm$  0.25 &   0.24  $\pm$  0.1 & 10.83  $\pm$  0.15 &  0.16  $\pm$  0.25 \\
        COS4\_27120 & 2.4780 & 56 $\pm$ 0.04 & 1.49  $\pm$  0.25 & -0.21  $\pm$  0.08 & 10.17  $\pm$  0.15 & -0.14  $\pm$  0.25 \\
        COS4\_27087 & 2.4794 & 45 $\pm$ 0.04 & 2.05  $\pm$  0.25 &  0.42  $\pm$  0.14 &  9.95  $\pm$  0.15 &  0.63  $\pm$  0.25 \\
        \enddata
\end{deluxetable*}

\subsection{Galaxy Properties}
We derive several galaxy properties such as SFR, $\Sigma_{\text{SFR}}$, M$_*$, $\Delta$MS, and inclination (Table \ref{tab:galaxy_properties}) to investigate how outflow velocity depends on these properties. 

SFR, $\Sigma_{\text{SFR}}$, M$_*$, and $\Delta$MS were presented previously in \cite{2019Schreiber}. SFR and M$_*$ were determined by modeling the broad- and medium-band spectral energy distributions (SEDs) spanning the optical to near-infrared range for each galaxy and supplemented with \textit{Spitzer} and \textit{Herschel} mid and far-IR photometry when available. SEDs were fit using \cite{2003Bruzual} population synthesis models that adopted the \cite{2000Calzetti} reddening law, solar metallicty, and SF histories. Furthermore, \cite{2019Schreiber} adopt a \cite{2003Chabrier} stellar initial mass function. There were 34 galaxies for which the SFR was determined by fitting SEDs across optical to \textit{Spitzer}/IRAC wavelengths. The remaining 73 galaxies had their SFR determined by combining the SFRs from \textit{Herschel/Spitzer} + UV. SFR and M$_*$ uncertainties were adopted from \cite{2018Tacconi}. An uncertainty of $\pm$0.25 dex is used for SED-inferred SFRs, while an uncertainty $\pm$0.2 dex is used for \textit{Herschel/Spitzer}-detected galaxies. We define the $\Sigma_{\text{SFR}}$ as:
\begin{equation}
    \Sigma_{\text{SFR}} = \frac{\text{SFR}}{2\pi R_e^2}
\end{equation}
The half-light effective radii ($R_e$) were obtained from \cite{2012ApJSvanderWel} and \cite{2014Lang}, who base their measurements on the H-band radii corrected to rest-frame 5000\r{A} using average color gradients. Uncertainties in $R_e$ were determined from GALFIT. We define the MS offset ($\Delta$MS) as:
\begin{equation}
    \Delta\text{MS} = \log(\text{SFR}/\text{SFR}\textsubscript{MS})
\end{equation}
where SFR$\textsubscript{MS}$ is the main sequence SFR for a given M$_*$ and redshift, as parameterized by \cite{2014Whitaker}. Uncertainties on $\Delta$MS are adopted to be the same as the uncertainties for the SFR.

Galaxy inclination ($i$) was calculated using the galaxy's axis ratio ($q$), where $q$ is the ratio of the minor to major axes. Neglecting the intrinsic thickness of the disk, we estimated $i = \arccos(q)$. \footnote{If one were to adopt a reasonable value for the finite intrinsic thickness (i.e., $\gamma=0.2$, where $\gamma$ is the ratio between the smallest and largest axes), at most, the difference would be $\sim7$ degrees for the most highly inclined systems in our sample.} Axis ratios were obtained from  \cite{2012ApJSvanderWel} based on the 3D-HST catalogs.

\subsection{AGN Identification}
From the sample of 80 galaxies used in our analysis, 12 are identified as hosting an AGN (15\%) on the basis of the narrow component [N\,{\sc ii}]$/\text{H}\alpha$ flux ratio and diagnostics from supplementary X-ray to mid-IR and radio data. Galaxies are identified as having an AGN when [N\,{\sc ii}]$/\text{H}\alpha\textsubscript{narrow} > 0.45$, or when characteristics indicative of an AGN are found in the radio, mid-IR, or X-ray \citep{2019Schreiber}. High ionization lines, such as N\,{\sc v}, Si\,{\sc iv}, and C\,{\sc iv}, that are indicative of AGN are also detected in the LRIS spectra. We find that 9 of the 12 galaxies identified as AGN based on characteristics from multi-wavelength data also have signatures of AGN in their LRIS spectra. The remaining 3 do not have emission of high ionization lines. In addition, we find 2 galaxies have AGN signatures in the LRIS spectra that were not previously identified as hosting AGN, yielding a total sample of 14 galaxies with AGN signatures. We have confirmed that the trends described in Section~\ref{sec:Results} are unaffected if we either include or exclude objects identified as AGN.

\subsection{Composite Spectra}
Out of our total of 80 usable spectra, there are only 57 that have at least one rest-UV feature measured in LRIS spectra. Limiting our sample to galaxies where only absorption lines or only Ly$\alpha$ are measured may bias our results. To fold the full sample of 80 usable LRIS spectra, regardless of requiring significant detections of LIS or Ly$\alpha$ lines, we construct composite spectra of equal-number bins (26 or 27 galaxies per bin) according to different galactic properties. We use these composite spectra to evaluate the average outflow velocities within different bins of galaxy properties. To create the composite spectra, we shifted each individual flux-calibrated galaxy spectrum into the rest frame, interpolated the spectra onto a common wavelength grid, and calculated the median flux of the full bin at each wavelength. We measured $\Delta v_{\text{LIS}}$ and $\Delta v_{\text{Ly}\alpha}$ using the same methods as for individual spectra as described in Section \ref{sec:Outflow velocities}.

\section{Results}\label{sec:Results}
In this section, we search for relations between outflow velocity and various galactic properties (i.e., inclination, SFR, $\Sigma\textsubscript{SFR}$, M$_*$, and $\Delta\text{MS}$). We also analyze the relationship between these galactic properties and LRIS and KMOS outflow detection fractions. We use galaxies presented in Table \ref{tab:outflow_detections} for our analysis.

\subsection{Inclination}
\subsubsection{Individual Measurements}
In the nearby universe, it has been shown that galaxy outflows are collimated perpendicular to the disk, while inflows occur along the major axis of the galaxy \citep{1990Heckman, 2010Chen, 2012bNewman, 2019Concas,2019RobertsBorsani}. Using 140,625 galaxies from SDSS with $0.05 \leq z \leq 0.18$, \cite{2010Chen} find that the outflow velocity is greater for more face-on galaxies, demonstrating that, in the local universe, galactic outflows are collimated. Furthermore, \cite{2011Bordoloi} investigate the Mg\, {\sc ii} absorption strength of low-redshift ($0.5 < z < 0.9$) galaxies and find that Mg\, {\sc ii} absorption is associated with bipolar regions aligned with the disk axis. This suggests that the model for collimated outflows holds true up to $z \sim 1$. \cite{2012Kornei} study 72 star-forming galaxies at $z \sim 1$ and find that face-on galaxies with lower inclination exhibit faster outflows compared to more edge-on galaxies with higher inclination. These results suggest that galactic winds also appear collimated for galaxies at $z \sim 1$. Similarly, \cite{2014Rubin} analyze 105 galaxies at $0.3 \leq z \leq 1.4$ and find that the outflow detection rate depends on inclination. They find that outflows are detected in $\sim89\%$ of face-on galaxies ($i<30^\circ$) while outflows are only detected in $\sim45\%$ of edge-on galaxies ($i>50^\circ$). Contrary to these well established findings in the local universe, the situation at higher redshift is less clear \citep{2012Law,2012bNewman,2019Schreiber,2022Weldon}. Most relevant to this analysis, in the larger KMOS$\textsuperscript{3D}$ parent sample, \cite{2019Schreiber} found no significant link between the frequency of outflow detection and axis ratio (\textit{q}) in their sample of galaxies with $z\sim0.6-2.7$. 

 As shown in Figure \ref{fig:delvLIS_vs_i_KMOSLRIS} and Table \ref{tab:p_values}, we find that there is no correlation between outflow velocity and inclination in our sample. Furthermore, we find that there is no relationship between the outflow detection rates of either KMOS or LRIS and inclination. However, galaxies that exhibit significant outflows from $\Delta v_{\text{LIS}}$  with both KMOS and LRIS appear to tend more toward higher inclinations, meaning more edge-on galaxies.

\subsubsection{Inclination Stacks}
\begin{figure*}
  \centering
   \includegraphics[width=0.415\linewidth]{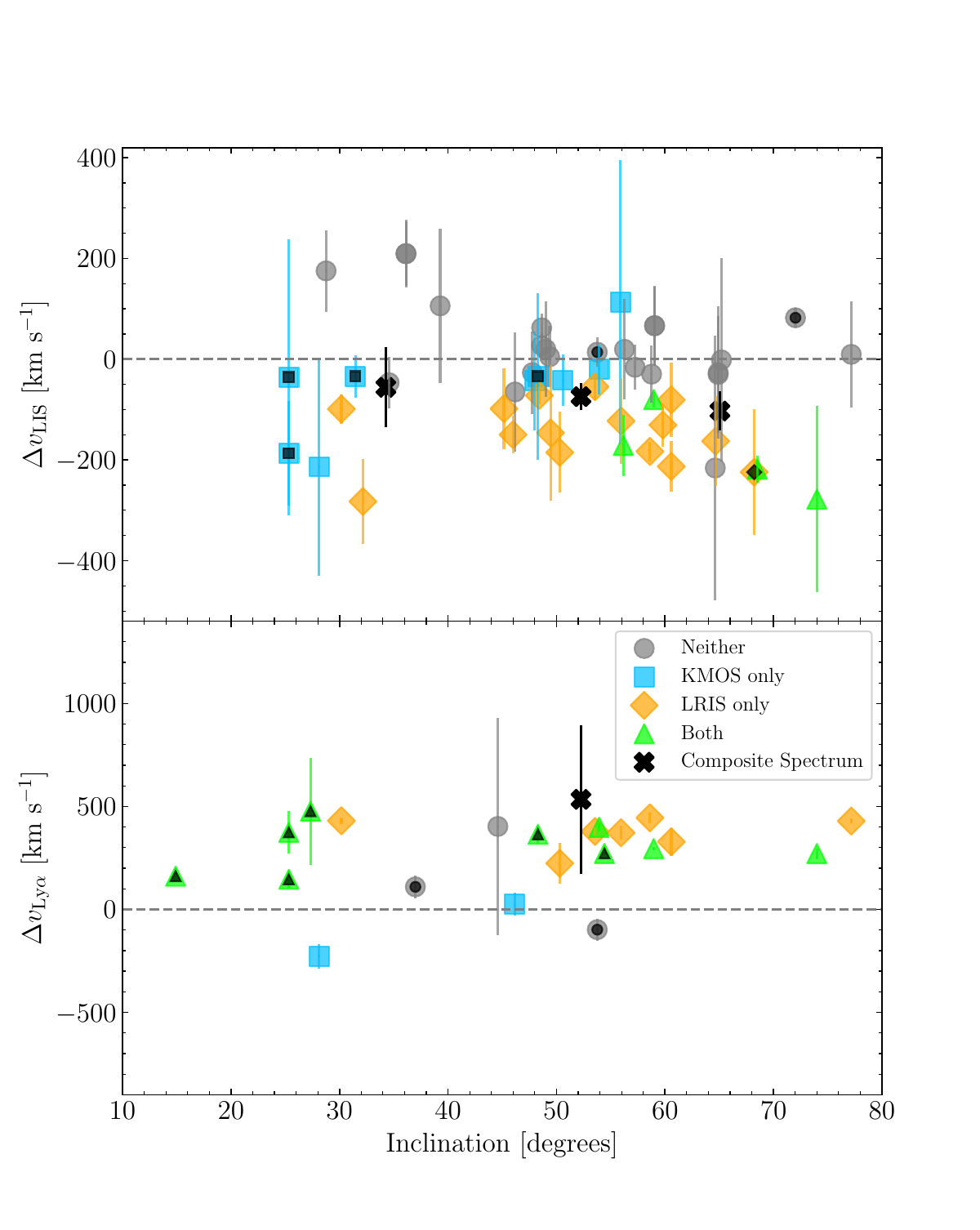}
    \includegraphics[width=0.535\linewidth]{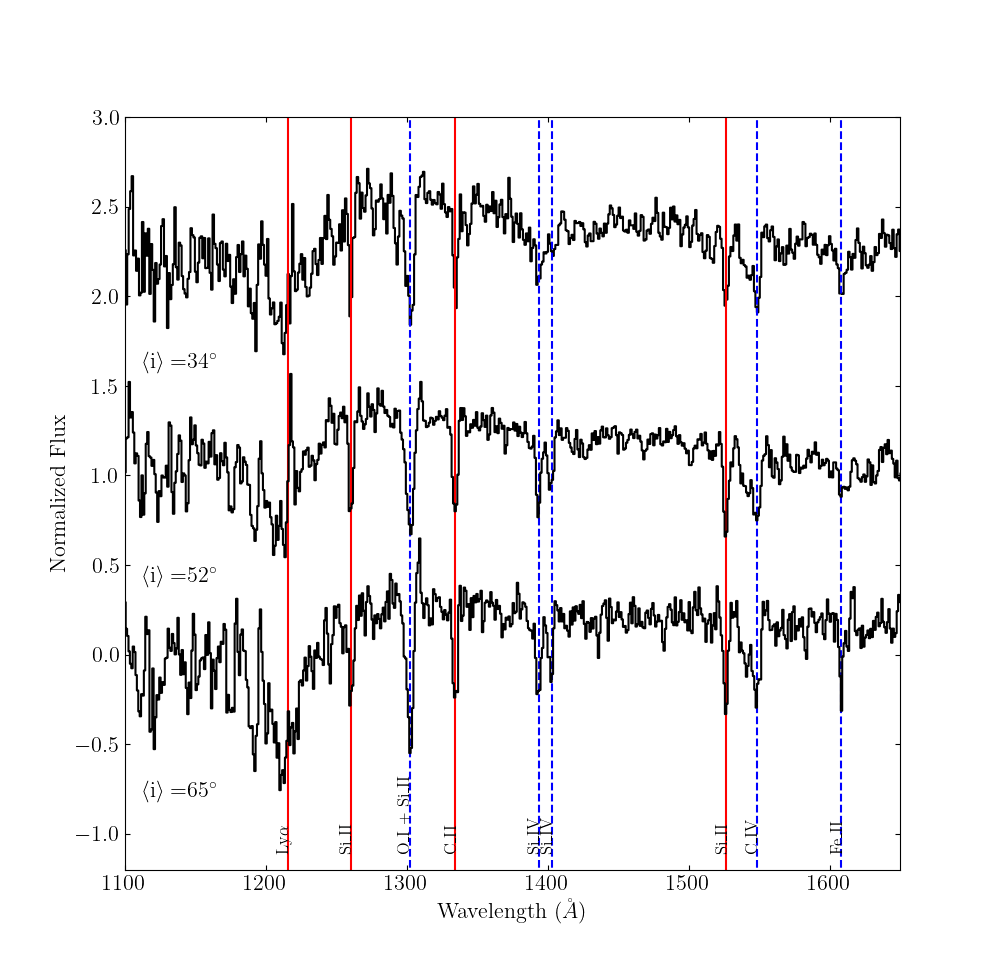}
  \caption{\textit{Left:} $\Delta v_{\text{LIS}}$ versus inclination (top) and $\Delta v_{\text{Ly}\alpha}$ versus inclination (bottom). Blue squares represent galaxies that had a significant outflow detection in KMOS, orange diamonds are for galaxies with a 1$\sigma$ outflow detection in the LRIS sample, gray circles are galaxies that had no significant outflow detections, and green triangles are for galaxies where both KMOS and LRIS found significant outflows. Black "X" symbols represent the composite spectra. In the bottom panel, there is only one composite spectrum data point, as Ly$\alpha$ emission was only detected in one of the three composite spectra. Data points with black markers inside indicate galaxies identified as hosting AGN. \textit{Right:} Composite spectra for three inclination bins. The top is the composite spectrum for our low inclination sample with $\langle i \rangle = 34^{\circ}$ (27 galaxies). The middle is the composite spectrum for our mid inclination sample with $\langle i \rangle =52^{\circ}$ (27 galaxies). The bottom is the composite spectrum for our high inclination sample with $\langle i \rangle = 65^{\circ}$ (26 galaxies). The laboratory wavelengths for Ly$\alpha$ and the absorption lines are plotted as vertical lines. Red lines indicate the lines used for our analysis while blue dotted lines are other features present in the spectra.}
    \label{fig:delvLIS_vs_i_KMOSLRIS}
\end{figure*}

To further analyze how outflow velocity depends on galaxy inclination, we split our sample into three equal-number bins based on inclination. The composite spectra, in order of increasing inclination, i, have $\langle i\rangle = 34^{\circ}, 52,^{\circ} 65^{\circ}$ with 27, 27, and 26 galaxies in each stack respectively (Figure \ref{fig:delvLIS_vs_i_KMOSLRIS}).  As shown in Figure \ref{fig:delvLIS_vs_i_KMOSLRIS}, we find no correlation between galaxy inclination and $\Delta v_{\text{LIS}}$. 

\subsection{Galaxy Stellar Properties}

\subsubsection{SFR}
A galaxy's SFR provides information on the amount of mechanical energy and radiation pressure available for driving outflows in star-forming galaxies. Several studies have found that outflow velocities increase with increasing SFR \citep{2005Martin, 2005Rupke, 2009Weiner, 2014Bordoloi, 2015Chisholm, 2017Sugahara,2021Prusinski}. The relation between outflow velocity and SFR can provide information on the driving mechanisms of the outflows. Specifically, if outflow velocity weakly depends on SFR, the outflow velocity may be driven by mechanical energy from supernovae or stellar winds \citep{2000Heckman, 2006Ferrara, 2010Chen}. If the outflow velocity is strongly dependent on SFR, the outflow velocity may be radiatively driven \citep{2012Sharma}. Many other studies have failed to find such a correlation due to a limited range in SFR available in their data \citep{2010Steidel, 2012Kornei, 2012bNewman}. \cite{2022Weldon} probe 155 galaxies with SFRs spanning $2-93 M_\odot \text{yr}^{-1}$, and find an absence of correlation between $\Delta v_{\text{LIS}}$ and SFR. This suggests that $\Delta v_{\text{LIS}}$ is potentially influenced by the presence of stationary gas near the systemic redshift of the galaxy. Furthermore, they find there is a small correlation between $\Delta v_{\text{Ly}\alpha}$ and SFR, which indicates that galactic outflows are driven by radiation pressure or supernova \citep{1985Chevalier,2011Murray}. 

\begin{figure*}
  \centering
   \includegraphics[width=0.4\linewidth]{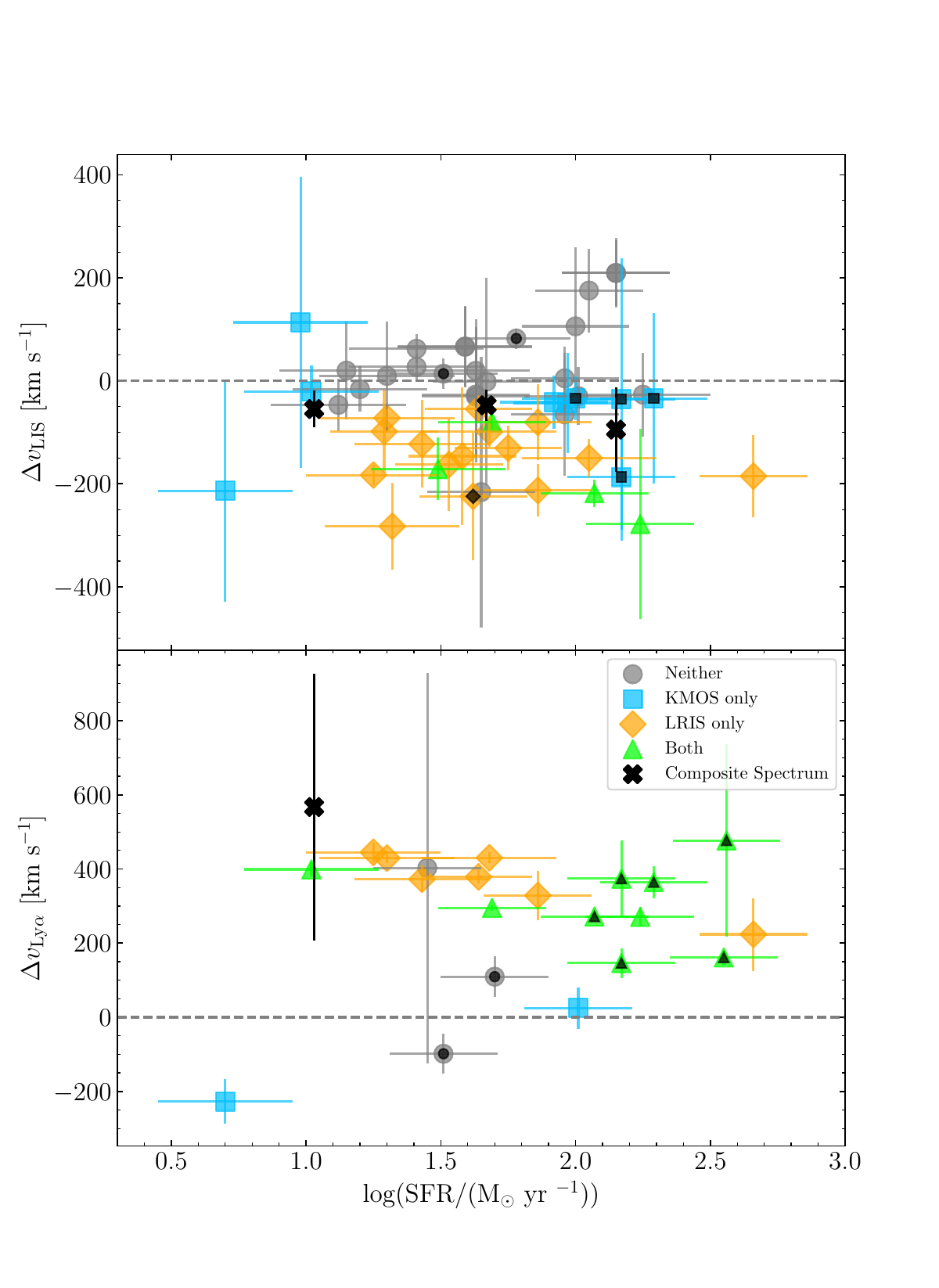}
    \includegraphics[width=0.55\linewidth]{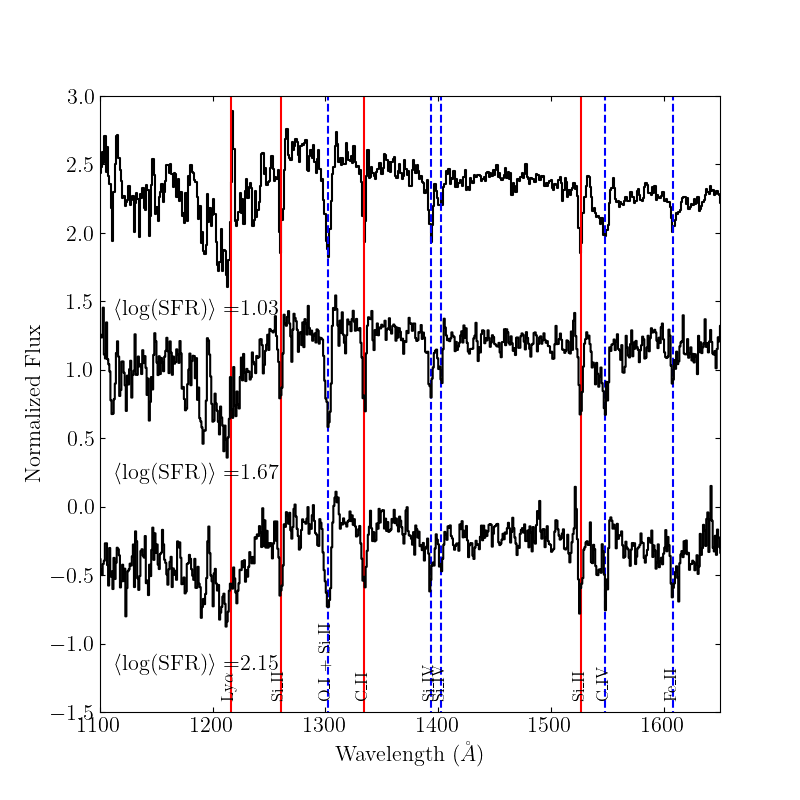}
  \caption{\textit{Left:} Same as Figure \ref{fig:delvLIS_vs_i_KMOSLRIS} but for $\Delta v\textsubscript{LIS}$ vs log(SFR) (top) and $\Delta v_{\text{Ly}\alpha}$ vs log(SFR) (bottom). In the bottom panel, there is only one composite spectrum data point, as Ly$\alpha$ emission was only detected in one of the three composite spectra. \textit{Right:} The same as Figure \ref{fig:delvLIS_vs_i_KMOSLRIS}, but for three log(SFR) bins. The top is the composite spectrum for our low inclination sample with $\langle$log(SFR)$\rangle = 1.03$ (27 galaxies). The middle is the composite spectrum for our mid inclination sample with $\langle$log(SFR)$\rangle = 1.67$ (27 galaxies). The bottom is the composite spectrum for our high inclination sample with $\langle$log(SFR)$\rangle = 2.15$ (26 galaxies).}
    \label{fig:del_v_vs_SFR}
\end{figure*}

We find that there is no correlation between $\Delta v_\textsubscript{LIS}$ and SFR in our sample as shown in Figure \ref{fig:del_v_vs_SFR} and Table \ref{tab:p_values}. Furthermore, our composite spectra (Figure \ref{fig:del_v_vs_SFR}) show no trends between $\Delta v_\textsubscript{LIS}$ and SFR.

\begin{deluxetable*}{cccc}
\tabletypesize{\small}
\tablecaption{Correlation coefficients between outflow and galaxy properties \label{tab:p_values}} 
\tablehead{
\colhead{Galaxy Property} & \colhead{$\rho_{\Delta v\textsubscript{LIS}}$} & \colhead{$\rho_{\Delta v\textsubscript{Ly$\alpha$}}$} & \colhead{$\rho _{v\textsubscript{max}}$}
}
\startdata
i  &  $-0.17 (0.245)$  &  $0.26 (0.261)$  &  $-0.04 (0.810)$  \\
$\log (\text{SFR})$   & $0.04 (0.784)$    & $0.15 (0.511)$         &    $-0.44 (0.002)$ \\
$\log (\Sigma\textsubscript{SFR})$ & $-0.1 (0.502)$            &  $0.20 (0.397)$   & $-0.15 (0.291)$ \\
$\log (\text{M}_{*})$ & $-0.07 (0.612)$          &   $-0.29 (0.200)$          &   $-0.17 (0.234)$ \\
$\Delta \text{MS}$     &   $0.13 (0.349)$         & $0.44 (0.044)$   & $-0.27 (0.056)$ 
\\
\enddata
\tablenotetext{*}{Each column lists the correlation coefficient $\rho$ (with corresponding $p$-values in parentheses), between $\Delta v\textsubscript{LIS}$, $\Delta v \textsubscript{Ly$\alpha$}$, or $v\textsubscript{max}$ and one of several stellar properties.}
\end{deluxetable*}

\begin{deluxetable*}{cccccc}
\tabletypesize{\small}
\tablecaption{Mean values for several star properties for different outflow detection methods \label{tab:Detections_stellar_prop}}
\tablehead{
\colhead{Outflow Detection} & \colhead{$\langle \log (\text{SFR/M}_\odot \text{ yr}^{-1}) \rangle$} & \colhead{$\langle \log (\Sigma\textsubscript{SFR}/\text{M}_\odot \text{yr}^{-1} \text{kpc}^{-2}) \rangle$} & \colhead{$\langle \log (\text{M}_*/\text{M}_\odot) \rangle$} & \colhead{$\langle \Delta \text{MS} \rangle$} & \colhead{$\langle \text{i} \rangle$ (deg)}
}
\startdata
Neither   & $1.14 \pm 0.10$           & $-0.26 \pm 0.08$                  & $10.35 \pm 0.04$           & $-0.20 \pm 0.10$ & $51 \pm 2.9$                \\
KMOS Only & $1.91 \pm 0.11$            & $0.10 \pm 0.14$                  & $10.74 \pm 0.09$           & $0.03 \pm 0.10$ & $50 \pm 2.6$                  \\
LRIS Only & $1.63 \pm 0.08$            & $0.06 \pm 0.11$                   & $10.23 \pm 0.08$           & $0.06 \pm 0.08$ & $52 \pm 2.8$               \\
Both      & $2.03 \pm 0.13$            & $0.82 \pm 0.23$                   & $10.59 \pm 0.14$           & $0.12 \pm 0.07$ & $46 \pm 5.7$            \\ 
\enddata
\end{deluxetable*}

\subsubsection{SFR Surface Density}
Environments with elevated $\Sigma\textsubscript{SFR}$ have a higher surface density of radiation pressure, and the radiation pressure acting on dust grains is more efficient. Therefore, areas with high $\Sigma\textsubscript{SFR}$ may be more efficient at transporting momentum and energy from overlapping supernovae or stellar winds from massive stars into the ISM \citep{2005Veileux}. The combination of a higher concentration of star formation, meaning more radiation and higher density of SNe, along with efficient radiative coupling, from high concentrations of dust, results in conditions susceptible to launching outflows in high $\Sigma_{\text{SFR}}$ environments. In order for galaxies to sustain outflows, \cite{2001Heckman} proposed that galaxies must exceed a $\Sigma\textsubscript{SFR}$ threshold of $\sim 0.05 \text{ M}_\odot \text{ yr}^{-1} \text{ kpc}^{-2}$ for the \cite{2003Chabrier} IMF.

At $z \leq 1$, there is a relationship between $\Sigma\textsubscript{SFR}$ and outflow velocity \citep{2010Chen, 2012Kornei, 2014Rubin, 2015Chisholm}. At higher redshift ($z > 1$), \cite{2012bNewman} found that the relative strengths of the broad outflow and narrow star formation components in rest-optical (H$\alpha$) line emission showed the strongest difference with $\Sigma\textsubscript{SFR}$ among galaxy properties (at $20 \sigma$ between the low and high $\Sigma\textsubscript{SFR}$ bins); a finer parameter space sampling showed a steep increase around $\Sigma\textsubscript{SFR}$ of $\sim 1 \text{ M}_\odot \text{ yr}^{-1} \text{ kpc}^{-2}$, which could reflect the thicker, more turbulent gas-rich disks at earlier epochs.  In the large sample analyzed by \cite{2019Schreiber}, the incidence of star formation-driven outflows showed a smoother increase with the fraction exceeding 15\% at $\Sigma\textsubscript{SFR}$ of $\sim 1 \text{ M}_\odot \text{ yr}^{-1} \text{ kpc}^{-2}$. \cite{2019Davies} exploited the high resolution of the subset of SINS/zC-SINF sample observed with adaptive optics to investigate trends of broad outflow emission in H$\alpha + $[N\,{\sc ii}] by stacking spectra of spaxels ($\sim 1 - 2$ kpc scales) in bins of local physical properties across all 28 non-AGN galaxies, finding a consistent but somewhat lower threshold of $\Sigma_\mathrm{SFR} \sim 0.3 \mathrm{M}_\odot\mathrm{yr}^{-1}\mathrm{kpc}^{-2}$, and derived v$\textsubscript{out}\propto\Sigma\textsubscript{SFR}^{0.34}$, intermediate between the shallow power-law for energy-driven winds \citep{2010Chen} and steeper power-law for momentum-driven winds \citep[e.g.,][]{2011Murray}. In contrast, \cite{2010Steidel} and \cite{2022Weldon} reported no correlation between outflow velocity and $\Sigma\textsubscript{SFR}$. \cite{2022Weldon} suggest the absence of observed correlation may stem from challenges in pinpointing the actual location of the gas and its coupling to star formation activity. This discrepancy could be exacerbated by potential limitations in LRIS observations, and the relationship may remain elusive due to its weak nature within constrained dynamical ranges of $\Sigma\textsubscript{SFR}$.

\begin{figure*}
  \centering
   \includegraphics[width=.42\linewidth]{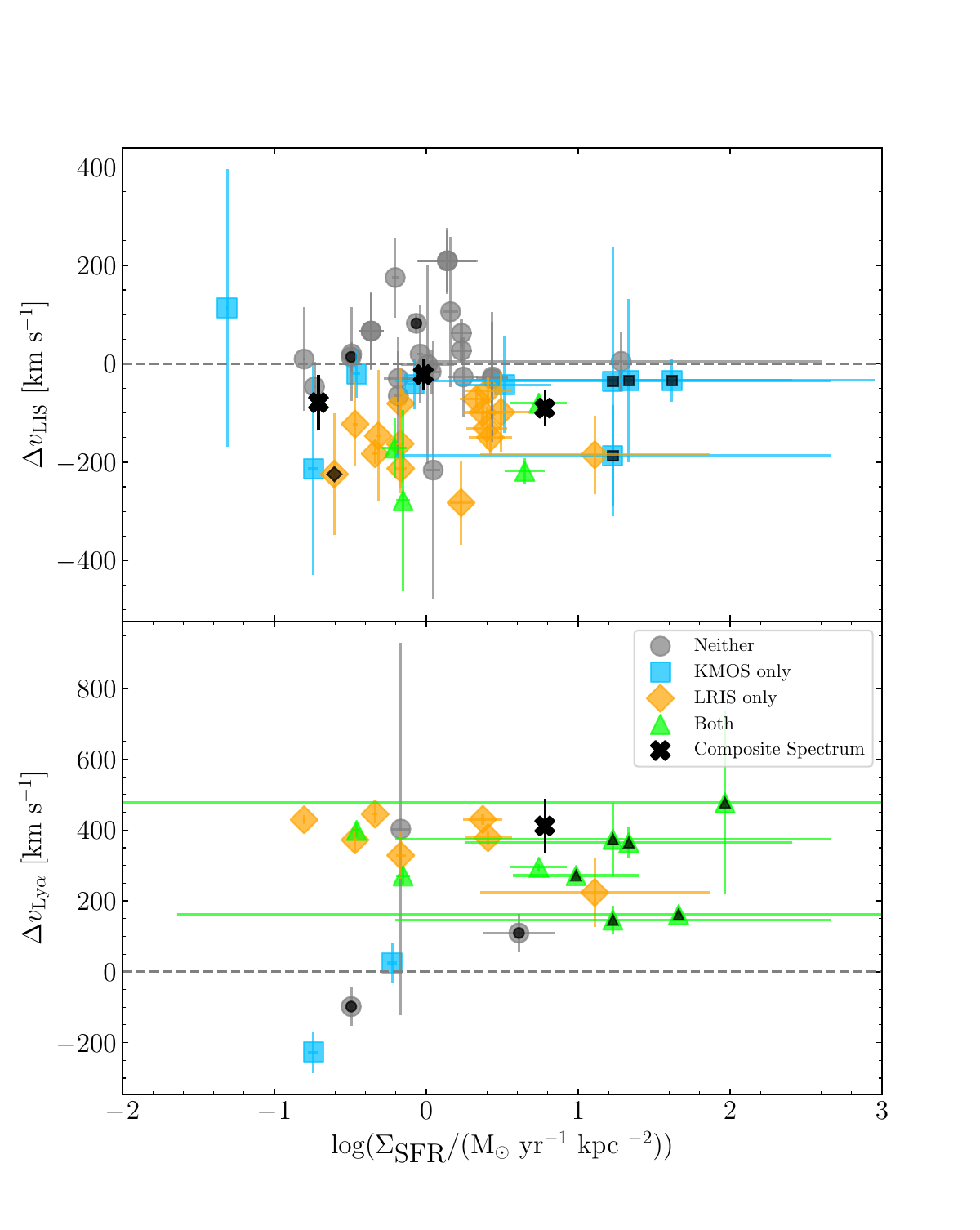}
    \includegraphics[width=0.53\linewidth]{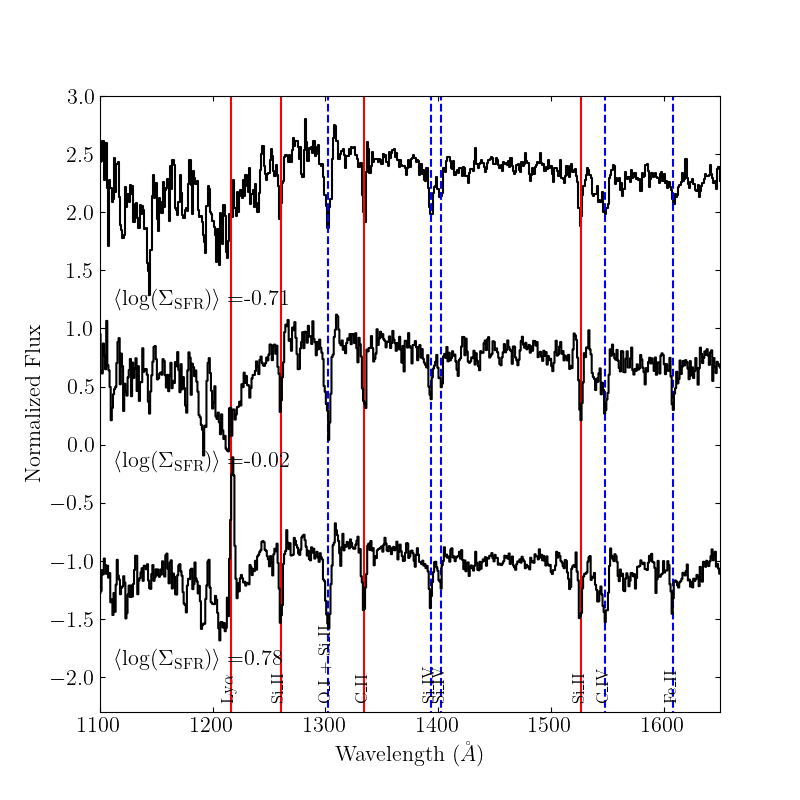}
  \caption{\textit{Left:} Same as Figure \ref{fig:delvLIS_vs_i_KMOSLRIS} but for $\Delta v\textsubscript{LIS}$ vs log($\Sigma\textsubscript{SFR}$) (top) and $\Delta v_{\text{Ly}\alpha}$ vs log($\Sigma\textsubscript{SFR}$ )(bottom). In the bottom panel, there is only one composite spectrum data point, as Ly$\alpha$ emission was only detected in one of the three composite spectra. \textit{Right:} The same as Figure \ref{fig:delvLIS_vs_i_KMOSLRIS}, but for three log($\Sigma\textsubscript{SFR}$) bins. The top is the composite spectrum for our low inclination sample with $\langle \log(\Sigma\textsubscript{SFR}) \rangle = -0.71$ (27 galaxies). The middle is the composite spectrum for our mid inclination sample with $\langle \log(\Sigma\textsubscript{SFR}) \rangle = -0.02$ (27 galaxies). The bottom is the composite spectrum for our high inclination sample with $\langle \log(\Sigma\textsubscript{SFR}) \rangle = 0.78$ (26 galaxies).}
  \label{fig:del_v_sigSFR}
\end{figure*}

As shown in Figure \ref{fig:del_v_sigSFR} and Table \ref{tab:p_values}, we find no significant trends between outflow velocity and $\Sigma\textsubscript{SFR}$ in our sample. Furthermore, our composite spectra (Figure \ref{fig:del_v_sigSFR}) also show no trends between outflow velocity and $\Sigma\textsubscript{SFR}$. 

\subsubsection{Stellar Mass}
Galaxies with a lower stellar mass (M$_*$) have a lower gravitational potential, resulting in a more efficient launch of outflows \citep{2022Reddy}. Our KMOS$\textsuperscript{3D}$ parent sample from \cite{2019Schreiber}, show that SF-driven winds show no significant dependence on stellar mass. For galaxies with a stellar mass at $\log(M_*/M_\odot) \geq 10.3$, SF-driven winds may not escape the galaxy but instead contribute to driving fountains \citep{1986Dekel, 2005Murray, 2008Oppenheimer, 2014Ubler}. \cite{2012Martin}, found that the detection rate of outflows does not rely on stellar mass. Additionally, \cite{2016Heckman} and \cite{2021Prusinski} find no correlation between stellar mass and outflow velocity. \cite{2019Schreiber} also find that the incidence, strength, and velocity of AGN-driven outflows are dependent on stellar mass, with most AGN-driven outflows detected above $\log(M_*/M_\odot) = 10.7$. 

\begin{figure*}
  \centering
   \includegraphics[width=.42\linewidth]{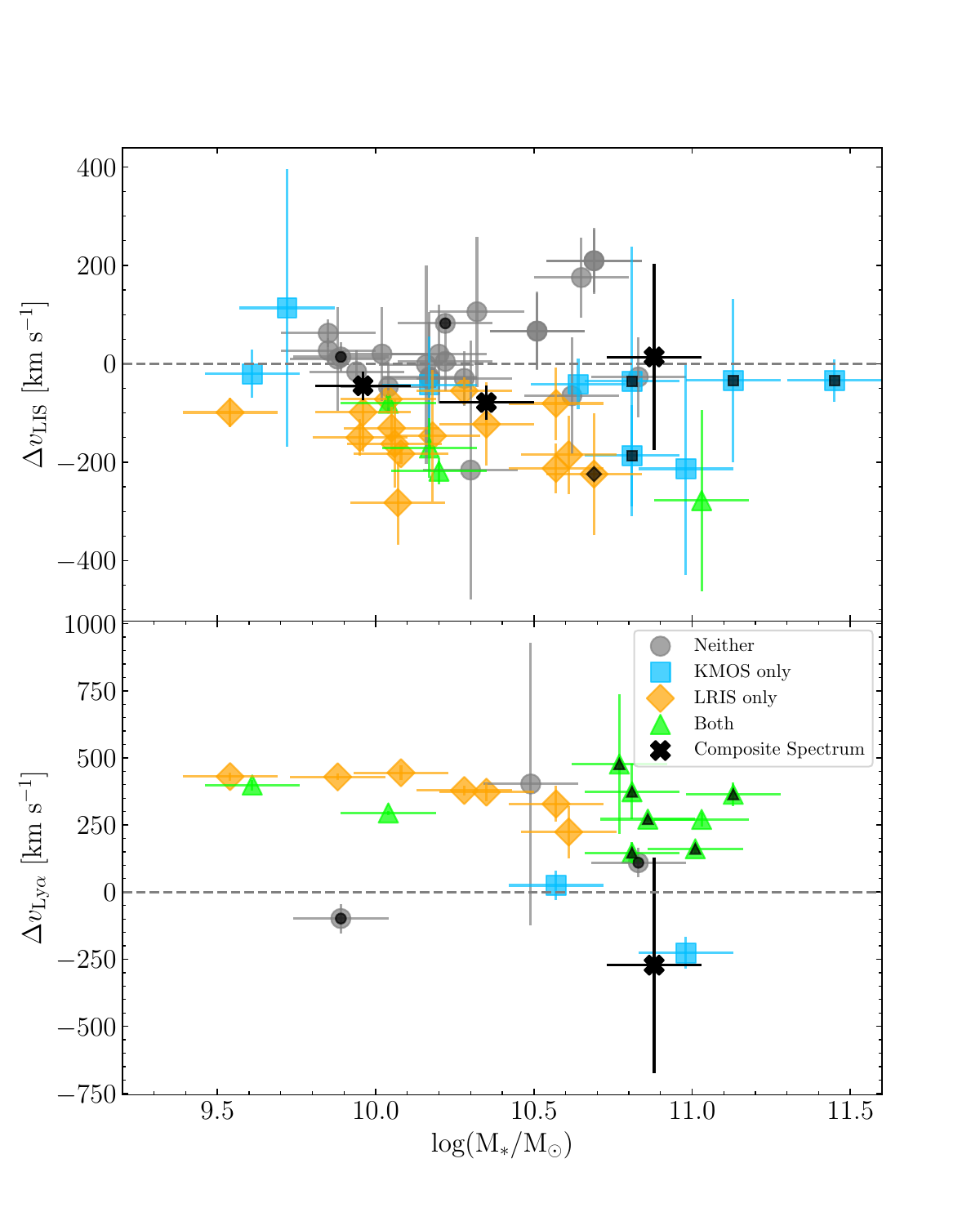}
    \includegraphics[width=0.53\linewidth]{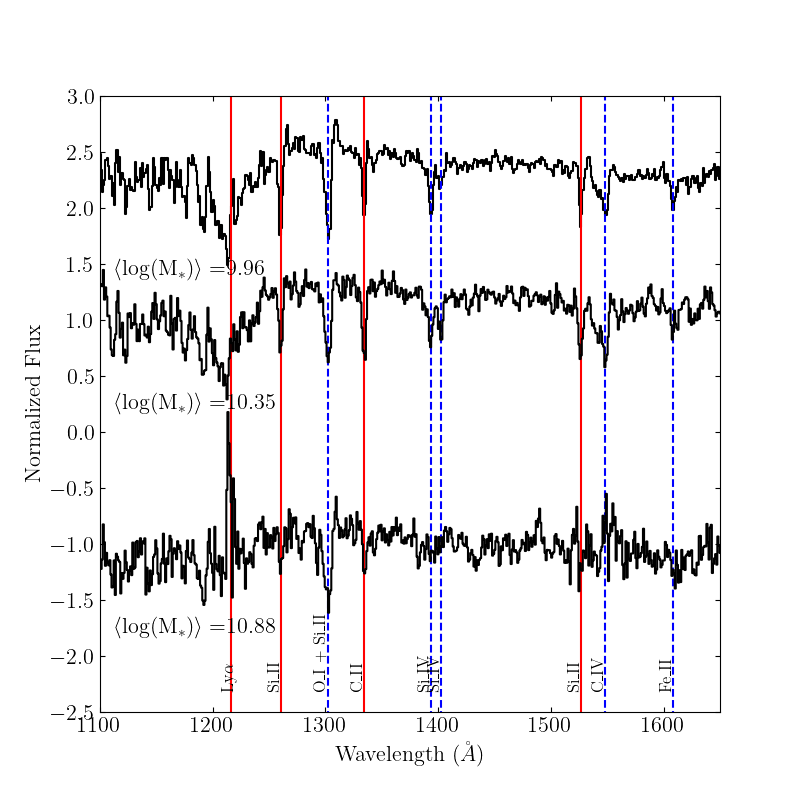}
  \caption{\textit{Left:} Same as Figure \ref{fig:delvLIS_vs_i_KMOSLRIS} but for $\Delta v\textsubscript{LIS}$ vs log(M$_*$) (top) and $\Delta v_{\text{Ly}\alpha}$ vs log(M$_*$) (bottom). In the bottom panel, there is only one composite spectrum data point, as Ly$\alpha$ emission was only detected in one of the three composite spectra. \textit{Right:} The same as Figure \ref{fig:delvLIS_vs_i_KMOSLRIS}, but for three M$_*$ bins. The top is the composite spectrum for our low inclination sample with $\langle$M$_*$$\rangle = 9.96$ (27 galaxies). The middle is the composite spectrum for our mid inclination sample with $\langle$M$_*$$\rangle = 10.35$ (27 galaxies). The bottom is the composite spectrum for our high inclination sample with $\langle$M$_*$$\rangle = 10.88$ (26 galaxies).}
  \label{fig:del_v_lMstar}
\end{figure*}

Drawing from the same KMOS$\textsuperscript{3D}$ sample as \cite{2019Schreiber}, we find that there is no correlation between outflow velocity and M$_*$ (Figure \ref{fig:del_v_lMstar} and Table \ref{tab:p_values}). Figure \ref{fig:del_v_lMstar} also illustrates that the composite spectra show no correlation as well. Furthermore, we find no correlation between outflow velocity and the method in which the outflow was detected (Table \ref{tab:Detections_stellar_prop}).

\subsubsection{$\Delta$MS}
Using the KMOS$\textsuperscript{3D}$ sample, \cite{2019Schreiber} find that AGN-driven outflows are not correlated with $\Delta$MS while SF-driven outflows are detected at higher $\Delta$MS. At $\Delta\text{MS}\geq +0.6$ dex, they find the highest percentage ($25-30\%$) of detected SF-outflows. These “starbursting outliers” drive a SF-driven outflow that is detectable in the rest-optical line emission \citep{2011Rodighiero,2019Schreiber}.

\begin{figure*}
  \centering
   \includegraphics[width=.415\linewidth]{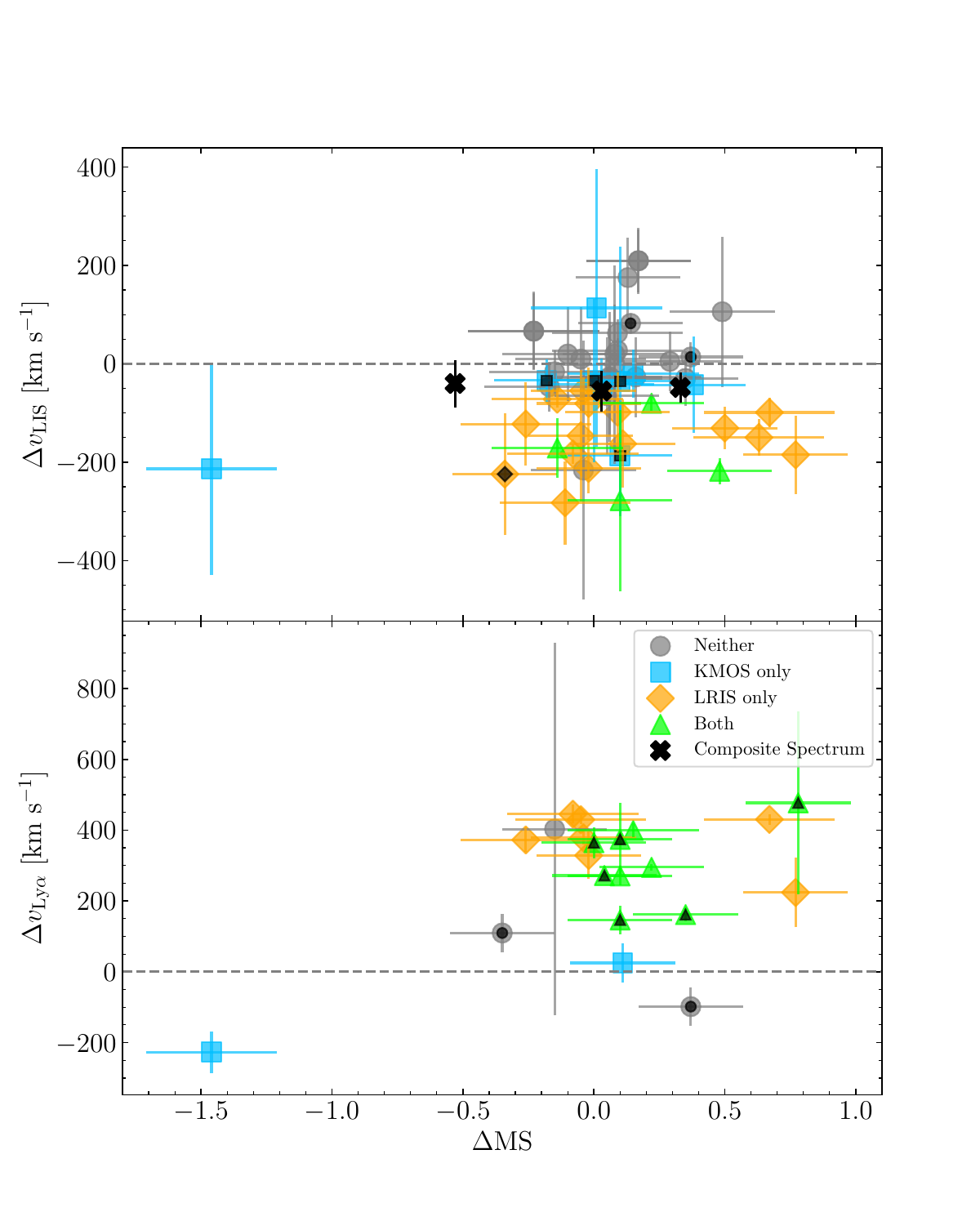}
    \includegraphics[width=0.535\linewidth]{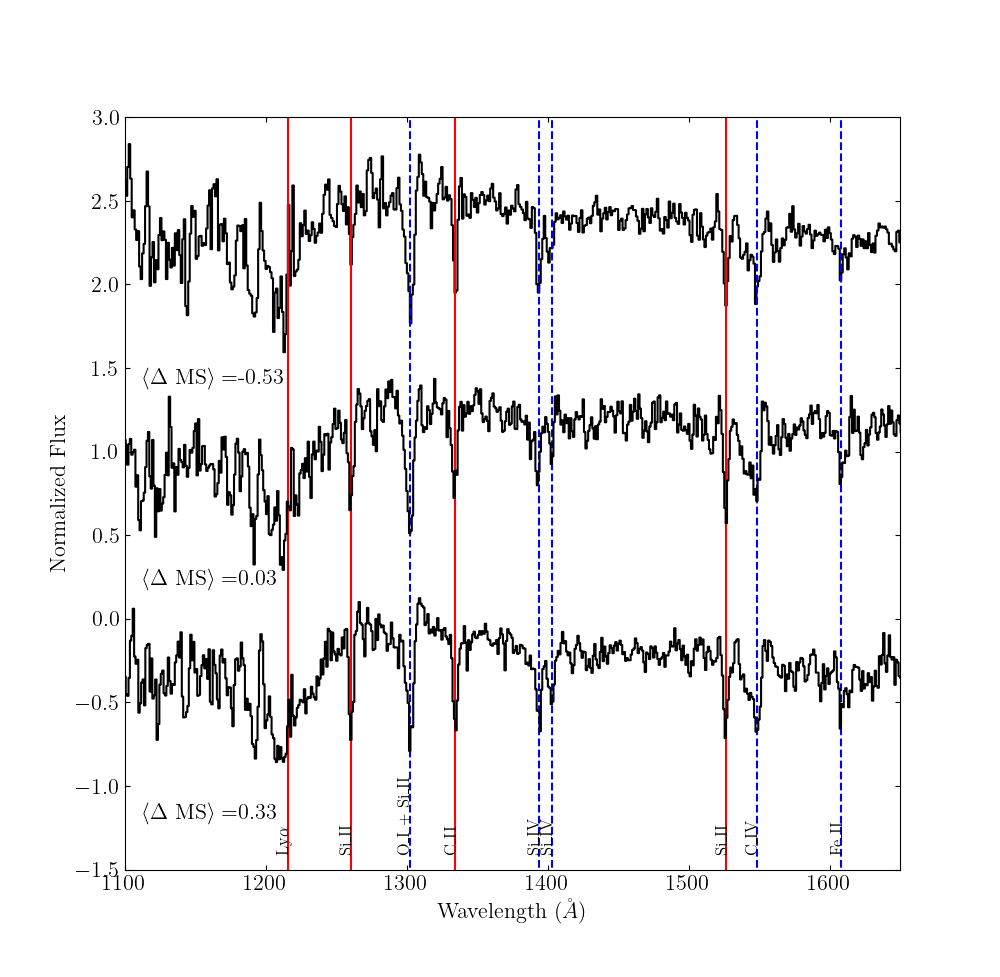}
  \caption{\textit{Left:} Same as Figure \ref{fig:delvLIS_vs_i_KMOSLRIS} but for $\Delta v\textsubscript{LIS}$ vs $\Delta$MS (top) and $\Delta v_{\text{Ly}\alpha}$ vs $\Delta$MS (bottom). In the bottom panel, there are no composite spectra data points, as Ly$\alpha$ emission was not detected in any of the three composite spectra. \textit{Right:} The same as Figure \ref{fig:delvLIS_vs_i_KMOSLRIS}, but for three $\Delta\text{MS}$ bins. The top is the composite spectrum for our low inclination sample with $\langle\Delta\text{MS}\rangle = -0.53$ (27 galaxies). The middle is the composite spectrum for our mid inclination sample with $\langle\Delta\text{MS}\rangle = 0.03$ (27 galaxies). The bottom is the composite spectrum for our high inclination sample with $\langle\Delta\text{MS}\rangle = 0.33$ (26 galaxies).}
  \label{fig:del_v_dMS}
\end{figure*}

As shown in Figure \ref{fig:del_v_dMS} and Table \ref{tab:p_values}, there is no correlation between $\Delta v \textsubscript{LIS}$ and $\Delta$MS among galaxies in our sample. While Table \ref{tab:p_values} suggests there may be a relationship between $\Delta v\textsubscript{Ly$\alpha$}$ and $\Delta$MS, ($p = 0.044$), the sample size is small with only 12 galaxies being included. Therefore, a larger sample is needed to robustly probe this relation. Figure \ref{fig:del_v_dMS} also illustrates that the composite spectra show no correlation.

\subsection{Maximum Outflow Velocity}
\begin{figure*}
    \centering
    \includegraphics[width = \linewidth]{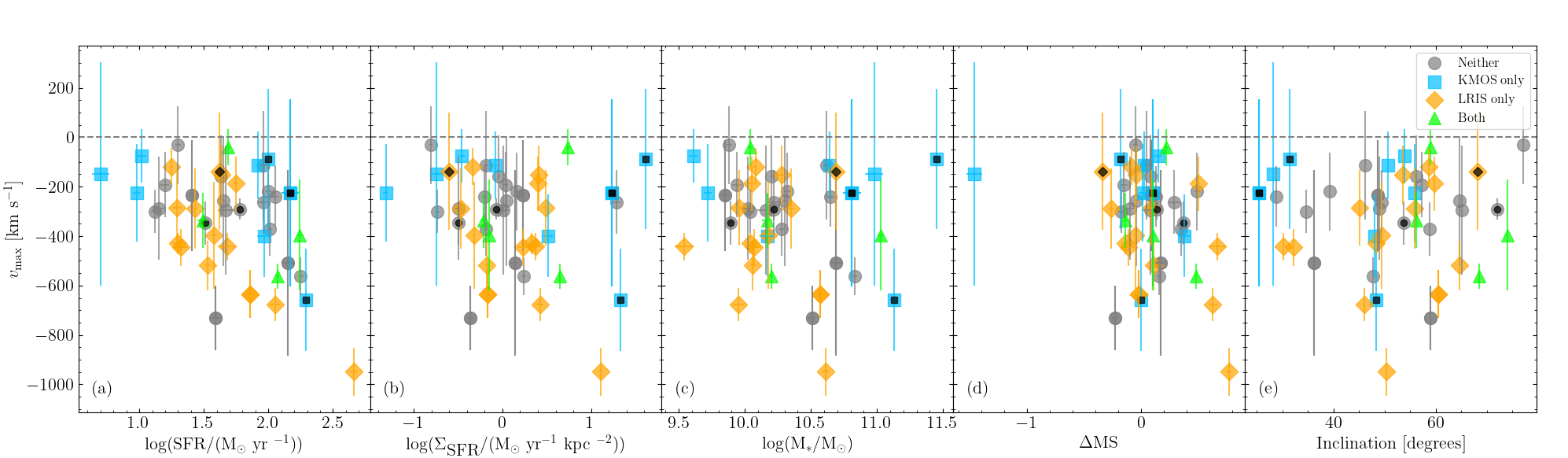}
    \caption{Relationships between $v\textsubscript{max}$ and the various galaxy properties. Grey circles indicate galaxies without significant outflow detections, blue squares indicate galaxies with a significant outflow detection in KMOS, orange diamonds indicate galaxies with a 1$\sigma$ outflow detection in the LRIS sample, and green triangles indicate galaxies where significant outflows were detected in both KMOS and LRIS. Relationships for \textit{(a):} SFR; \textit{(b):} $\Sigma\textsubscript{SFR}$; \textit{(c):}  M$_*$; \textit{(d):} $\Delta$MS; \textit{(e):} inclination.}
    \label{fig:vmax_stellar_prop}
\end{figure*}

In addition to analyzing the relationships between $\Delta v\textsubscript{LIS}$ and the various stellar properties, we test if the stellar properties are correlated with $v\textsubscript{max}$. As shown in Figure \ref{fig:vmax_stellar_prop} and Table \ref{tab:p_values}, our results for inclination, $\Sigma\textsubscript{SFR}$, M$_*$, and $\Delta$MS are qualitatively unchanged compared to our $v\textsubscript{LIS}$ results. However, we find a statistically significant correlation between $v\textsubscript{max}$ and SFR. Our results show that galaxies with a higher SFR have a higher maximum outflow velocity. These results agree with \cite{2022Weldon} and support the idea that supernova or radiation pressure drive galactic outflows \citep{1985Chevalier, 2011Murray}. Given this strong correlation, $v\textsubscript{max}$ might be a more reliable measure of the correlations between outflow velocity and other galactic stellar properties.

\subsection{Average Galaxy Properties of Outflow Samples}
\begin{figure*}
    \centering
    \includegraphics[width = \linewidth]{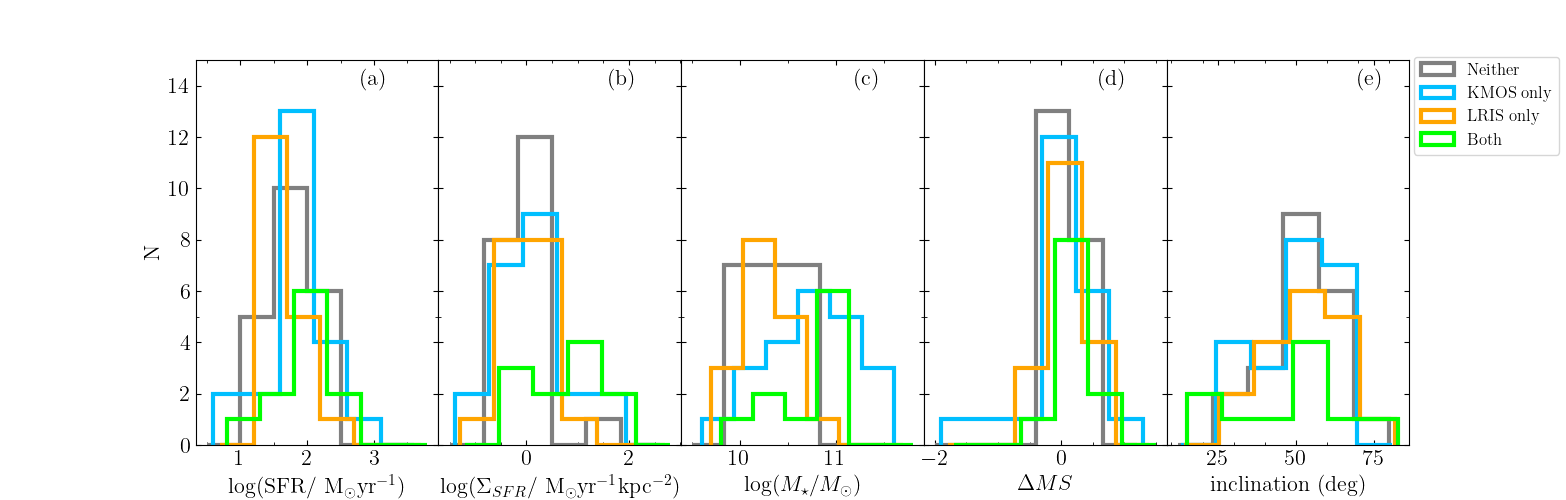}
    \caption{Distributions of stellar properties with different outflow detections. The blue bars represent galaxies with outflows detected with only KMOS \citep{2019Schreiber}, orange bars represent galaxies with outflows detected with LRIS only, and green bars represent galaxies with outflows detected from both KMOS and LRIS. Distributions of \textit{(a):} SFR; \textit{(b):} $\Sigma\textsubscript{SFR}$; \textit{(c):} M$_*$; \textit{(d):} $\Delta$MS; \textit{(e):} inclination.} 
    \label{fig:stellar_prop_hist}
\end{figure*}

While we do not observe strong correlations between outflow and galaxies properties in the scatter plots or stacked spectra, we do find a relationship within the distinct stellar population properties (SFR, $\Sigma\textsubscript{SFR}$, and $\Delta$MS) when looking at the stellar property distributions of the different outflow detections (Figure \ref{fig:stellar_prop_hist} and Table \ref{tab:Detections_stellar_prop}). We find that the average SFR, $\Sigma\textsubscript{SFR}$, and $\Delta$MS are significantly higher in cases where outflows are detected than when they are not detected. Table \ref{tab:Detections_stellar_prop} shows that the highest averages are for cases when outflows are detected in both KMOS and LRIS for SFR, $\Sigma\textsubscript{SFR}$, and $\Delta$MS. This result provides evidence that there is a connection between outflow and galaxy properties given that the mean star formation properties of the outflow samples are higher than the mean properties of the non-outflow samples. Figures \ref{fig:del_v_vs_SFR}, \ref{fig:del_v_sigSFR}, \ref{fig:del_v_dMS} show no significant trends with outflow speed, suggesting that these correlations are noisy.

\begin{figure*}
    \centering
    \includegraphics[width=\linewidth]{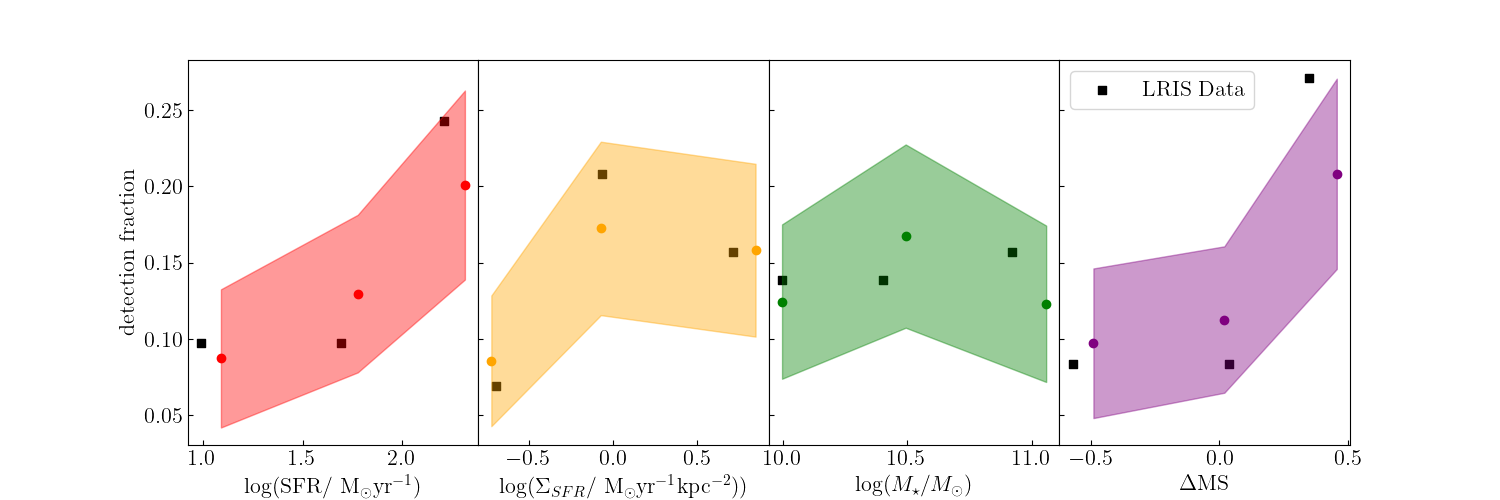}
    \caption{Results from 1000 realizations of entire KMOS-LRIS-size samples drawn from the entire KMOS$\textsuperscript{3D}$ sample. Each simulation drew 80 galaxies at $z = 1.4-2.7$ from the 600 galaxy parent sample and separated the galaxies into 3 bins based on the stellar properties. The bins contain 27, 27, and 26 galaxies, respectively. The colored data points represent the average values found from the simulations, the shaded regions represent the uncertainty, and the black squares are the averages found from our LRIS sample.}
    \label{fig:jackknife_exp}
\end{figure*}

To test these correlations, we performed jackknife simulations by pulling 80 random galaxies from the parent KMOS$\textsuperscript{3D}$ sample within the same redshift range as our LRIS sample to create 1000 mock samples. We split these mock samples into 3 bins based on the stellar properties, and found the average detection fraction for each bin. Our simulations recover trends consistent with our LRIS sample with large uncertainties (Figure \ref{fig:jackknife_exp}), suggesting that a larger sample (i.e, a sample twice as large) spanning a wider dynamic range in galaxy properties is required.

Higher SFR, $\Sigma\textsubscript{SFR}$, and $\Delta$MS in galaxies with detected outflows also suggest that outflows may only be launched and detectable above some threshold of SFR, $\Sigma\textsubscript{SFR}$, and $\Delta$MS \citep{1977McKee, 2002Heckman, 2014Rubin, 2015Chisholm, 2020RobertsBorsani, 2022Weldon}. This is possible if the maximum outflow velocity that can be produced from star formation alone has an Eddington limit \citep{2005Murray, 2005Thompson, 2010Hopkins}, or if the outflow speed is modulated by the density of the surrounding ISM.

\section{Discussion} \label{sec:Discussion}

\subsection{Outflow Kinematics}
As shown in Figure \ref{fig:Delta_v_hist}, we find that $\langle \Delta v_{\text{LIS}}\rangle= -56 \pm 16 \text{ km s}^{-1}$ and $\langle \Delta v_{\text{Ly}\alpha} \rangle =  266 \pm 41 \text{km s}^{-1}$. \cite{2010Steidel} use a similar sample to ours (i.e., 89 galaxies at $2 \leq z \leq 3$) to quantify a galaxy outflow velocities using interstellar absorption lines (IS) (C\,{\sc ii} $\lambda 1334$, Si\,{\sc iv} $\lambda 1393$, Si\,{\sc ii} $\lambda 1526$) and Ly$\alpha$. These authors find  $\langle \Delta v_{\text{IS}}\rangle= -164 \pm 16 \text{km s}^{-1}$ and $\langle \Delta v_{\text{Ly}\alpha} \rangle =  445 \pm 27 \text{km s}^{-1}$, which are significantly more offset from zero than our results.

Although \cite{2010Steidel} do not report error bars on outflow velocities, we can look at the general shape of the $\Delta v\textsubscript{LIS}$ and $\Delta v_{\text{Ly}\alpha}$ histograms to analyze the differences in our average outflow velocities. \cite{2010Steidel} find 11 out of their 89 galaxies to have $\Delta v\textsubscript{IS} \geq 0$. Moreover, all of their $\Delta v\textsubscript{IS} \leq +100 \text{ km s}^{-1}$. We find 15 out of 80 galaxies in our sample have $\Delta v\textsubscript{LIS} \geq 0$ with 5 galaxies having $\Delta v\textsubscript{LIS} \geq 100 \text{ km s}^{-1}$. Furthermore, \cite{2010Steidel} find no Ly$\alpha$ measurements that are bluer than $100 \text{ km s}^{-1}$, whereas we find 3 galaxies measurements in the $v_{\text{Ly}\alpha} < 0$ regime (Figure \ref{fig:Delta_v_hist}). 

One source for these discrepancies may be the sample selection criteria. Although both \cite{2010Steidel} and our sample investigate main sequence star-forming galaxies, we select our samples differently. Most notably, \cite{2010Steidel} select galaxies in the rest-UV down to a fixed rest UV luminosity with a median absolute UV magnitude of $M_{UV}=-20.42$, while our parent sample, KMOS$\textsuperscript{3D}$ from \cite{2019Schreiber}, was selected down to fixed rest-optical luminosity with a median absolute UV magnitude of $M_{UV}=-19.93$, 0.5~magnitudes fainter in the rest-UV. These differences sample UV continuum luminosity selection may translate into differences in the properties of gas flows probed, but a detailed sample comparison is outside the scope of this work. 

We find that our results are in agreement with other work that show lower velocities than those reported in \cite{2010Steidel}. For example, \cite{2022Weldon} use the MOSDEF-LRIS survey and measured $\Delta v\textsubscript{LIS}$ and $\Delta v_{\text{Ly}\alpha}$ for 155 star forming galaxies at $1.42 \leq z \leq 3.48$. A small overlap exists in our samples since we include 22 galaxies from the MOSDEF-LRIS survey, yet the samples are largely independent. In the MOSDEF-LRIS sample, they found that the peak distribution for $\Delta v\textsubscript{LIS}$ was $-60$ km s$^{-1}$. Furthermore, as in our sample, there are several galaxies with $\Delta v_{\text{Ly}\alpha} < 100$ km s$^{-1}$. \cite{2012Talia} and \cite{2022Calabro} also present results consistent with those presented here. \cite{2012Talia} used 74 rest-frame UV spectra from the Galaxy Mass Assembly ultra-deep Spectroscopic Survey (GMASS) with a redshift range of $1.5 \leq z \leq 2.8$. They used composite spectra to calculate the velocities of the strongest interstellar absorption lines (a mixture of low and high ionization lines denoted as $\Delta v\textsubscript{IS}$) and found the average $\Delta v\textsubscript{IS} \sim -100$ km s$^{-1}$. \cite{2022Calabro} study 330 galaxies at $2 \leq z \leq 4.6$ using the VANDELS survey and find the average velocity traced by UV absorption lines was $\Delta v\textsubscript{IS} = -60 \pm 10$ km s$^{-1} $. Furthermore, they report a positive velocity shift for 39\% of their sample. 

These results present outflows with velocities closer to zero, suggesting a more nuanced picture \citep{2010Steidel}. The smaller velocities may imply that there is greater absorption at the galaxies systemic redshift that is more prominent in $z\sim2$ galaxies. A more thorough decomposition of the absorption line may be necessary, separating the component associated with the galaxy disk from the one associated with an outflow \citep{2009Weiner, 2010Steidel, 2012Martin, 2014Rubin}. Furthermore, we find that the Ly$\alpha$ kinematics are more complex given that there is evidence of blueshifted gas in the Ly$\alpha$, which is indicative of infalling gas \citep{2006Verhamme,2012Kulas, 2023Weldon}.

\subsection{Inferred Geometry}
Our results show no strong correlation with inclination implying the outflows are not collimated (Figure \ref{fig:delvLIS_vs_i_KMOSLRIS}). Outflows with a spherical geometry and unity covering fraction would have a $\sim$ 100\% detection rate. However, we find outflows detected with KMOS have a 30\% detection rate while outflows detected with LRIS have a 49\% detection rate. KMOS outflows are detected with H$\alpha$ kinematics while LRIS outflow detections are found with LIS absorption lines and Ly$\alpha$ emission. The lower detection rate for KMOS H$\alpha$ outflows implies that H$\alpha$ may be more sparsely distributed with lower covering fraction while neutral outflow gas traced from LIS absorption lines covers a larger solid angle surrounding the galaxy. Moreover, these tracers might exhibit sensitivity to varying timescales of outflow activity. For instance, H$\alpha$ provides a more instantaneous insight as it explores material closer to the launching site of outflows. In contrast, rest-UV lines could be dispersed across greater distances and lower densities along the line of sight. In future work, we will analyze the geometry of outflows more closely with JWST images in the COSMOS and GOODS-S fields. Furthermore, as in previous work (e.g., \cite{2010Steidel, 2022Weldon}), we find that the connections between outflow and galaxy properties are noisy with our small dynamic range. A larger sample is needed with a wide enough dynamic range to robustly probe these relations. In addition, larger samples resolved on sub-galactic scales (e.g., expanding previous studies with AO-assisted IFU of star-forming galaxies and/or strongly lensed sources), will aid by enabling a more direct association of the local stellar properties and outflow launching sites and potential provide better constraints on outflow geometry.

\section{Conclusions}
We utilized a sample of 80 galaxies with a redshift range of $1.4 \leq z \leq 2.7$ to investigate galaxy outflows. To explore the multi-phase nature of galaxy outflows, we use a novel data set that includes both LRIS and KMOS in order to probe galaxy outflows in both H$\alpha$ and the rest-UV. Outflows are identified in galaxies by using broad H$\alpha$ ($+$[N\,{\sc ii}]$+$[S\,{\sc ii}]) emission or by identifying low ionization interstellar absorption lines or Ly$\alpha$ emission. Outflow velocities are measured from rest-UV features. We also examine how outflow velocity depends on various galactic properties such as SFR, $\Sigma_{\text{SFR}}$, M$_*$, $\Delta$MS, and inclination. Our key results are as follows:
\begin{enumerate}
  \item The mean velocities of our sample are $\langle \Delta v_{\text{LIS}}\rangle= -56 \pm 16 \text{ km s}^{-1}$ and $\langle \Delta v_{\text{Ly}\alpha} \rangle =  266 \pm 41 \text{ km s}^{-1}$. These average velocities, lower than those found in previous work at a similar redshift range, suggest that the interstellar absorption lines have a multi-component structure (i.e., one component from the galaxy disk and one component from the galaxy outflow).
  \item In our sample, we find no significant correlation between outflow velocity from absorption line centroids and galaxy properties.  However, we find that the average SFR, $\Sigma\textsubscript{SFR}$, and $\Delta$MS are significantly higher in galaxies where outflows are detected, reflecting underlying trends in incidence of outflow detection reported for larger samples, such as the parent sample of 599 SFGs \citep{2019Schreiber}. In addition, we find that quantifying outflow kinematics in terms of the maximum outflow velocity, $v\textsubscript{max}$, may be more sensitive to underlying correlations.
  \item Outflow velocity is not correlated with inclination, implying that outflows are not collimated. Furthermore, we did not have a 100\% detection rate meaning outflows cannot be spherical either. These two results suggest that outflows are sparsely distributed. We find that outflows detected with H$\alpha$ have a 30\% detection rate while galaxies detected with LIS absorption lines have a 49\% detection rate meaning that LIS absorption lines cover a larger solid angle. Furthermore, LIS absorption lines trace longer scales and lower densities along the line of sight of outflow activity compared to H$\alpha$.
\end{enumerate}

We find that the correlations between outflow properties and galaxy properties have a significant amount of intrinsic scatter. Thus, a larger sample with a wider dynamic range and a a sample that explores these correlations on spatially-resolved scales are needed to better understand these relationships. Furthermore, higher resolution rest-optical imaging from JWST will enable a more robust exploration of the geometry of galactic outflows. A full analysis of such observations will be crucial for a full understanding of outflows.

\begin{acknowledgments}
We acknowledge support from NSF AAG grants 2009313 and 2009085.
We also wish to extend special thanks to those of Hawaiian ancestry on
whose sacred mountain we are privileged to be guests. Without their generous
hospitality, the work presented herein would not have been possible.
\end{acknowledgments}

%


\bibliography{ms}{}
\bibliographystyle{aasjournal}



\end{document}